\documentclass[11pt,a4paper,column]{article}
\usepackage[left=2cm,right=2cm,top=2cm,bottom=2cm]{geometry}
\usepackage{graphicx}
\usepackage{dcolumn}
\usepackage{bm}
\usepackage{array}
\usepackage{booktabs}
\usepackage{multirow}
\newcommand{\head}[1]{\textnormal{\textbf{#1}}}
\newcommand{\normal}[1]{\multicolumn{1}{l}{#1}}
\begin{document}
\begin{center}
\begingroup
    \fontsize{16pt}{16pt}\selectfont
  Solution of QCD$\otimes$QED coupled DGLAP equations at NLO \\\ \\\
\endgroup
\end{center}

\begin{center}
\begingroup
    \fontsize{14pt}{12pt}\selectfont
   S. Zarrin and G.R. Boroun
\endgroup
\end{center}
\begin{center}
\footnotesize{\begingroup  \fontsize{11pt}{11pt}\selectfont Physics Department, Razi University, Kermanshah 67149, Iran \endgroup}
\\\
\end{center}
   \begin{center}
    Corresponding author:\\\
    G.R. Boroun\\\
    Physics Department\\\
Razi University\\\
Kermanshah 67149\\\
        Iran\\\
E-mail: grboroun@gmail.com, boroun@razi.ac.ir\\\ \\\ \\\
\end{center}
\begingroup
\begin{center}
\fontsize{11pt}{12pt}\selectfont
\textbf{Abstract}
\end{center}
\endgroup
\begingroup
\fontsize{11pt}{11pt}\selectfont
In this work, we present an analytical solution for QCD$\otimes$QED coupled Dokshitzer-Gribov-Lipatov-Altarelli-Parisi (DGLAP)
 evolution equations at the leading order (LO) accuracy in QED and next-to-leading order (NLO) accuracy in perturbative QCD using double Laplace transform. This technique
  is applied to obtain the singlet, gluon and photon distribution functions and also the proton
  structure function. We also obtain contribution of photon in proton at LO and NLO at high energy and successfully compare the proton structure function with HERA data \cite{12} and APFEL results \cite{7}. Some comparisons also have been done for the singlet and gluon distribution functions with the MSTW results\cite{9}. In addition, the contribution of photon distribution function inside the proton has been compared with results of MRST \cite{11} and with the contribution of sea quark distribution functions  which obtained by MSTW \cite{9} and CTEQ6M \cite{14}.
\endgroup \\\

\subsection*{I. Introduction}
In quantum electrodynamic (QED), interactions of photon can be
regarded as a structureless object since QED is an abelian gauge
theory and photon has no self-interaction. The photon in an
interaction can be fluctuate into a charged fermion and
anti-fermion, because of Heisenberg uncertainty principle, and if
one of fermions interacts with a gauge boson the photon reveals
its parton structure. Fig. (1) shows scheme of deep inelastic
scattering a photon with a gauge boson. In the deep inelastic
scattering of electron-positron collider LEP and the
electron-proton collider HERA are reported the main results on the
structure of photon. Recent studies on the effect of Drell-Yan
with high-mass in ATLAS have shown which the structure of photon
or corrections of QED have effects on parton distribution
functions \cite{1,21}. These results and discoveries improve our
understanding about the internal structure of the proton and it
can approximate theoretical activity to experimental data.

In Ref. \cite{11}, Martin and et al. showed that the photon
distribution is larger than the $b$ quark distribution at  $Q^{2}=
20 GeV^{2}$ and also larger than the sea quarks at the highest
values of $x$ inside the proton and neutron. So, it is interesting
to study the photon distribution of the proton and neutron, to
obtain these contributions at different scales can be used
QCD$\otimes$QED coupled DGLAP evolution equations. Recently,
several methods have been proposed to solve the coupled DGLAP
evolution equations as Laplace transform [8-12] and Mellin
transform methods \cite{3} and etc. The most appropriate and
simplest of these methods is the laplace transfom, because it
simplifies the equations to simplest form. Block et al. in Ref.
\cite{22} showed that the NLO coupled DGLAP evolution equations,
by using the double Laplace transform, can be solved and arrived
to decoupled NLO evolved solutions. In this method, the Laplace
transforms are respect to $x$ and $Q^{2}$ and these transforms
determine the singlet $F_{s}(x,Q^{2})$ and gluon $G(x,Q^{2})$
distribution functions directly, using as input $F^{s}_{0}(x)
\equiv F^{s}(x,Q^{2}_{0})$ and $G_{0}(x) \equiv G(x,Q^{2}_{0})$
where $Q^{2}_{0}$ is initial scale. According to  Ref. \cite{22},
it can be realized the accuracy of this method.
\begin{figure}[h]
\begin{center}
\includegraphics[width=.4\textwidth]{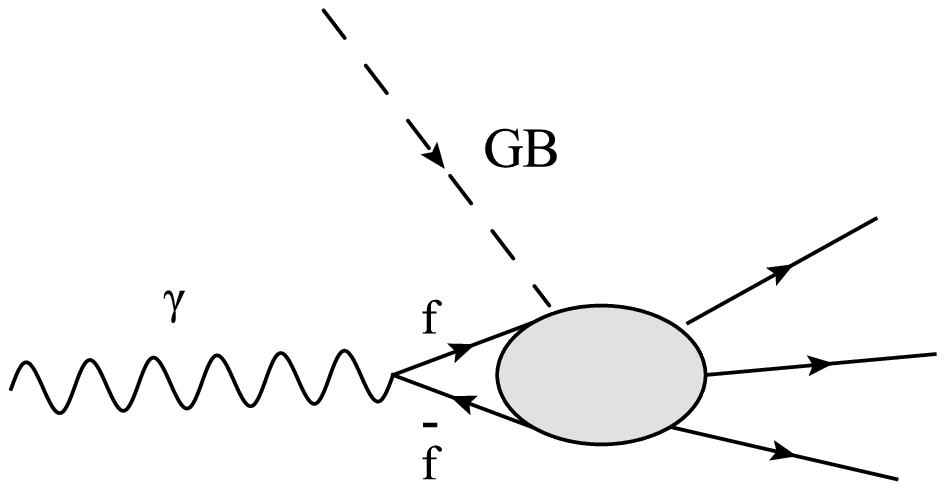}
\quad

\begingroup
    \fontsize{10pt}{12pt}\selectfont  {
\label*{
Figure 1: Probing the structure of a quasi-real photon by a gauge boson (GB) in deep inelastic scattering}}.
\endgroup
\end{center}
\end{figure}

In this work, we try to apply this method to solve of
QCD$\otimes$QED coupled DGLAP equations at LO and NLO QCD, since
QED contributions have effects on the proton structure functions.
Also, the individual singlet, gluon, photon distributions from
starting distributions are analyticaly calculated using the double
Laplace transform technique and extracted  the photon distribution
function in proton and  proton structure function. By this method
and QCD$\otimes$QED coupled DGLAP equations, it can be obtained
the singlet, gluon and photon distribution functions as follows
\begin{eqnarray}
  F_{2}^{s}(x,Q^{2})&=&\mathcal{F}(F^{s}_{0}(x),  G_{0}(x), M_{0}(x))\nonumber\\
 &&\mathrm{and}\nonumber\\
G(x,Q^{2})&=&\mathcal{G}(F^{s}_{0}(x),  G_{0}(x), M_{0}(x))\nonumber \\
 &&\mathrm{and}\nonumber\\
 M(x,Q^{2})&=&\mathcal{M}(F^{s}_{0}(x),  G_{0}(x), M_{0}(x)),\nonumber
\end{eqnarray}
 which $\mathcal{F}$, $\mathcal{G}$ and $\mathcal{M}$ are functions which can
 be obtained using splitting functions and  $ F^{s}(x,Q^{2}_{0})$, $G(x,Q^{2}_{0})$
 and $M(x,Q^{2}_{0})$, which are the singlet, gluon and photon distribution functions
 at initial scale respectively. By this method the singlet, gluon and photon distribution
 functions at arbitrary $Q^{2}$ can expressed as a convolution of these function at initial
 scale. The obtained results of the double Laplace transform show that the photon distribution
 function affect on value of singlet and gluon distribution functions inside the proton, however
 this effect is small,  but  these results show that the contribution of   photon distribution function
 is larger than contributions of $b$ quark at middle $Q^{2}$ and high $x$. This contribution is significant
 in analogy with the contributions of sea quarks at high $x$. And also, these results show  that  photon distribution
 function at  energy scales higher than initial scale depends on gluon distribution function at initial scale.

The remainder of this article is organized as follows: Section II
involves an general solution for the decoupling of QCD$\otimes$
QED DGLAP evolution equations analytically at LO
analyses with respect to the double Laplace transform method. In
Section III, we use this method to calculate DGLAP evolution equations at NLO in QCD and LO in QED. In Section IV the proton
distribution function have been obtained by Laplace transform, and some compression
are presented between our results and available HERA data \cite{12}  and
APFEL results \cite{7} . In this section, the results gluon and
singlet distribution compared with MSTW\cite{9} and at the end of
this section, we present most important part of this article,
which is the photon distribution function and show comparison this
distribution with sea quarks MSTW \cite{9} and CTEQ6M \cite{14}.
In the last, we give our conclusions. In Appendix, we present
results of a set of coefficients in Laplace $s$ space, which are
functions of splitting functions.

\subsection*{II. Master formula for LO corrections }
The QCD$\otimes$QED  coupled DGLAP equations at LO for the evolution of parton distribution function, quarks, antiquarks, gluon and photon expressed as follows \cite{11},
$$
\frac{\partial q(x,Q^{2})}{\partial \ln Q^{2}}=\frac{\alpha_{s}(Q^{2})}{2\pi}\left\lbrace P_{qq}(x)\otimes q_{i}(x,Q^{2})+P_{qg}\otimes g(x,Q^{2})\right\rbrace+
$$
\begin{equation}
\frac{\alpha_{e}(Q^{2})}{2\pi}\left\lbrace e_{i}^{2} \tilde{P}_{qq}(x)\otimes q_{i}(x,Q^{2})+e_{i}^{2}P_{q\gamma}(x)\otimes \gamma(x,Q^{2})\right\rbrace,
\end{equation}
$$
\frac{\partial \bar{q}(x,Q^{2})}{\partial \ln Q^{2}}=\frac{\alpha_{s}(Q^{2})}{2\pi}\left\lbrace P_{\bar{q}\bar{q}}(x)\otimes \bar{q}_{i}(x,Q^{2})+P_{\bar{q}g}\otimes g(x,Q^{2})\right\rbrace+
$$
\begin{equation}
\frac{\alpha_{e}(Q^{2})}{2\pi}\left\lbrace e_{i}^{2} \tilde{P}_{\bar{q}\bar{q}}(x)\otimes \bar{q}_{i}(x,Q^{2})+e_{i}^{2}P_{\bar{q}\gamma}(x)\otimes \gamma(x,Q^{2})\right\rbrace,
\end{equation}
\begin{equation}
\frac{\partial g(x,Q^{2})}{\partial \ln Q^{2}}=\frac{\alpha_{s}(Q^{2})}{2\pi}\left\lbrace P_{gq}(x)\otimes\sum_{i}\left(q_{i}(x,Q^{2})+\bar{q}(x,Q^{2})\right)+P_{gg}\otimes g(x,Q^{2})\right\rbrace
\end{equation}
\begin{equation}
\frac{\partial \gamma(x,Q^{2})}{\partial \ln Q^{2}}=\frac{\alpha_{e}(Q^{2})}{2\pi}\left\lbrace P_{\gamma q}(x)\otimes\sum_{i}e_{i}^{2}\left(q_{i}(x,Q^{2})+\bar{q}(x,Q^{2})\right)+P_{\gamma \gamma}\otimes \gamma(x,Q^{2})\right\rbrace,
\end{equation}
where the splitting functions $P_{i,j}(x)$ are the Altarelli-Parisi splitting kernels at one loop correction,  which are defined as follows,
$$
P_{qq}^{LO}=\frac{4}{3}\frac{1+x^{2}}{(1-x)_{+}}+2\delta(1-x) , \ \ \ P_{qg}^{LO}=\frac{1}{2}\left(x^{2}+(1-x)^{2}\right),
$$\\
$$
P_{gq}^{LO}=\frac{4}{3}\left(\frac{1+(1-x)^{2}}{x}\right),\ \ \ P_{gg}^{LO}=6\left(\frac{1-x}{x}+\frac{x}{(1-x)_{+}}+x(1-x)\right)+ \left(\frac{11}{2}-\frac{n_{f}}{3}\right)\delta(1-x),
$$\\
$$
P_{\bar{q}\bar{q}}^{LO}=P_{qq}^{LO},\ \ \ P_{\bar{q},g}^{LO}=P_{q,g}^{LO},\ \ \ \tilde{P}_{\bar{q}\bar{q}}^{LO}=\tilde{P}_{qq}^{LO}, \ \ \ P_{\bar{q} \gamma}^{LO}=P_{q \gamma}^{LO},
$$
\begin{equation}
\tilde{P}_{qq}^{LO}=C_{F}^{-1}P_{qq}^{LO},\ \ \ P_{\gamma q}^{LO}=C_{F}^{-1} P_{gq}^{LO}, \ \ \ P_{q \gamma}^{LO}=T_{R}^{-1}P_{qg}^{LO},\ \ \ P_{\gamma \gamma}^{LO}=-\frac{2}{3}\sum_{i}e_{i}^{2}\delta(1-x),
\end{equation}
where $C_{F}=\frac{4}{3}$, $T_{R}=\frac{1}{2}$, $n_{f}$ is the number of active quark
 flavours, QCD and QED running coupling constant at LO are as,
 $$\alpha_{s}^{LO}(Q^{2})=\frac{4\pi}{\beta_{0}\ln({\frac{Q^{2}}{\Lambda^{2}}})},  \\\\\\\  \alpha_{e}(Q^{2})=\frac{\alpha_{e} (\mu)}{1-\frac{\alpha_{e} (\mu)}{3\pi}\ln({\frac{Q^{2}}{\mu^{2}}})},$$
where  $\beta_{0}$ is the one loop (LO) correction to the QCD $\beta$-function and $\alpha_{e}(1eV)=1/137$. In the DGLAP evolution equations, we take $n_{f}=4$ and $\Lambda=192 MeV$ for $m_{c}^{2}<Q^{2}\leq m_{b}^{2}$ and $n_{f}=5$ and $\Lambda=146 MeV$ for $m_{b}^{2}<Q^{2}$, these values have been used in CTEQ5L \cite{4}. By summing Eqs.(1,2) and using $F^{p}_{2}=5/18F^{s}$ approximation, we can write,
$$
\frac{\partial F^{s}(x,Q^{2})}{\partial \ln Q^{2}}=\frac{\alpha_{s}^{LO}(Q^{2})}{2\pi}\left\lbrace P_{qq}^{LO}(x)\otimes F^{s}(x,Q^{2})+2n_{f}P_{qg}^{LO}\otimes G(x,Q^{2})\right\rbrace+
$$
\begin{equation}
\frac{\alpha_{e}(Q^{2})}{2\pi}\left\lbrace  \frac{5}{18}\tilde{P}_{qq}^{LO}(x)\otimes F^{s} (x,Q^{2})+2A P_{q\gamma}^{LO}(x)\otimes M(x,Q^{2})\right\rbrace,
\end{equation}
\begin{equation}
\frac{\partial G(x,Q^{2})}{\partial \ln Q^{2}}=\frac{\alpha_{s}^{LO}(Q^{2})}{2\pi}\left\lbrace P_{gq}^{LO}(x)\otimes F^{s}(x,Q^{2})+P_{gg}^{LO}\otimes G(x,Q^{2})\right\rbrace,
\end{equation}
\begin{equation}
\frac{\partial M(x,Q^{2})}{\partial \ln Q^{2}}=\frac{\alpha_{e}(Q^{2})}{2\pi}\left\lbrace \frac{5}{18} P_{\gamma q}^{LO}(x)\otimes F^{s} (x,Q^{2})+P_{\gamma \gamma}^{LO}\otimes M(x,Q^{2})\right\rbrace.
\end{equation}
To use of Laplace transform, we apply the variable change $x{\equiv}\exp(-\upsilon)$, $y{\equiv}\exp(-\omega)$,  and also we define the following functions,
 \begin{equation}
 \hat{F}^{s}(\upsilon,Q^{2}) {\equiv}
F^{s}(e^{-\upsilon},Q^{2}), \ \ \hat{G}(\upsilon) {\equiv}
G(e^{-\upsilon}),\ \ \hat{M}(\upsilon,Q^{2}) {\equiv}
M(e^{-\upsilon},Q^{2}).
\end{equation}
Defining the Laplace transform method, we have,
$$
 f(s,Q^{2})=\mathcal{L}\left[\hat{F}^{s}(\upsilon,Q^{2});s\right]=\int_{0}^{\infty}\hat{F}^{s}(\upsilon,
Q^{2})e^{-s\upsilon}d\upsilon
$$
 \begin{equation}
g(s,Q^{2})=\mathcal{L}\left[\hat{G}(\upsilon,Q^{2});s\right],\ \ m(s,Q^{2})=\mathcal{L}\left[\hat{M}(\upsilon,Q^{2});s\right].
\end{equation}
The Laplace transform converts Eqs. (6-8)  into three coupled ordinary first order differential equations in $s$ space and these can be written as
$$
\frac{\partial f}{\partial \ln Q^{2}}(s,Q^{2})=\frac{\alpha_{s}^{LO}(Q^{2})}{4\pi}\Phi_{f}^{LO}(s)f(s,Q^{2})+\frac{\alpha_{s}^{LO}(Q^{2})}{4\pi}\Theta_{f}^{LO}(s)g(s,Q^{2})+\frac{\alpha_{e}(Q^{2})}{4\pi}\Upsilon_{f}^{LO}(s)f(s,Q^{2})
$$
\begin{equation}
+\frac{\alpha_{e}(Q^{2})}{4\pi}\Omega_{f}^{LO}(s)m(s,Q^{2}),
 \end{equation}
 \begin{equation}
\frac{\partial g}{\partial \ln Q^{2}}(s,Q^{2})=\frac{\alpha_{s}^{LO}(Q^{2})}{4\pi}\Phi_{g}^{LO}(s)g(s,Q^{2})+\frac{\alpha_{s}^{LO}(Q^{2})}{4\pi}\Theta_{g}^{LO}(s)f(s,Q^{2}),
 \end{equation}
 \begin{equation}
\frac{\partial m}{\partial \ln Q^{2}}(s,Q^{2})=\frac{\alpha_{e}(Q^{2})}{4\pi}\Omega_{m}^{LO}(s)m(s,Q^{2})+\frac{\alpha_{e}(Q^{2})}{4\pi}\Upsilon_{m}^{LO}(s)f(s,Q^{2}),
 \end{equation}
where coefficients $\Phi^{LO}$, $\Psi^{LO}$, $\Omega^{LO}$, and $\Upsilon^{LO}$ are the leading-order splitting functions at Laplace $s$ space by;
\begin{equation}
\Phi_{f}^{LO}\left(s\right)=4-\frac{8}{3}\left(\frac{1}{s+1}+\frac{1}{s+2}+2\left(\psi(s+1)+\gamma_{E}\right)\right),
\end{equation}
\begin{equation}
\Theta_{f}^{LO}\left(s\right)=2n_{f}\left(\frac{1}{s+1}-\frac{2}{s+2}+\frac{2}{s+3}\right),
\end{equation}
\begin{equation}
\Upsilon_{f}^{LO}\left(s\right)=\frac{5}{24}\Phi_{f}^{LO}\left(s\right),\ \ \ \Omega_{f}^{LO}\left(s\right)=\frac{A}{n_{f}}\Theta_{f}^{LO}\left(s\right),
\end{equation}
\begin{equation}
\Phi_{g}^{LO}\left(s\right)=\frac{33-2n_{f}}{3}+12\left(\frac{1}{s}-\frac{2}{s+1}+\frac{1}{s+2}-\frac{1}{s+3}-\psi(s+1)-\gamma_{E}\right),
\end{equation}
\begin{equation}
\Theta_{g}^{LO}\left(s\right)=\frac{8}{3}\left(\frac{2}{s}-\frac{2}{s+1}+\frac{1}{s+2}\right),
\end{equation}
 \begin{equation}
\Omega_{m}^{LO}\left(s\right)=-\frac{4A}{3}, \ \ \ \Upsilon_{m}^{LO}\left(s\right)=\frac{5}{24}\Theta_{g}^{LO}\left(s\right) ,
 \end{equation}
indeed $\psi(s)$ is the digamma function,$\gamma_{E}$ is Euler's constant and $A=\sum_{i=1}^{n_{f}}e_{i}^{2}$. Introducing the new variable $\tau$ as $\frac{\partial \tau(Q^{2})}{\partial \ln(Q^{2})}=\frac{\alpha_{s}^{LO}(Q^{2})}{4\pi}$, the coupled first order differential equations in s-space
can be rewritten
 \begin{equation}
\frac{\partial f}{\partial \tau}(s,\tau)=\Phi_{f}^{LO}(s)f(s,\tau)+\Theta_{f}^{LO}(s)g(s,\tau)+\frac{\alpha_{e}(Q^{2})}{\alpha_{s}^{LO}(Q^{2})}\Upsilon_{f}^{LO}(s)f(s,\tau)+\frac{\alpha_{e}(Q^{2})}{\alpha_{s}^{LO}(Q^{2})}\Omega_{f}^{LO}(s)m(s,\tau),
 \end{equation}
 \begin{equation}
\frac{\partial g}{\partial \tau}(s,\tau)=\Phi_{g}^{LO}(s)g(s,\tau)+\Theta_{g}^{LO}(s)f(s,\tau),
 \end{equation}
 \begin{equation}
\frac{\partial m}{\partial \tau}(s,\tau)=\frac{\alpha_{e}(Q^{2})}{\alpha_{s}^{LO}(Q^{2})}\Omega_{m}^{LO}(s)l(s,\tau)+\frac{\alpha_{e}(Q^{2})}{\alpha_{s}^{LO}(Q^{2})}\Upsilon_{m}^{LO}(s)f(s,\tau).
 \end{equation}
Generally to do a calculation with high accuracy, we use the following expression for the  $\frac{\alpha_{e}(Q^{2})}{\alpha_{s}^{LO}(Q^{2})}$ as
\begin{equation}
\frac{\alpha_{e}(Q^{2})}{\alpha_{s}^{LO}(Q^{2})}\approx a_{10}+a_{11}\exp(b_{11}\tau),
\end{equation}
where the constants $a_{10}$, $a_{11}$, $b_{11}$ are found by fitting. To solve and decouple QCD $\otimes$ QED DGLAP evolutions, we need transform Eqs. (20-22)  from $\tau$ space to $u$ space by the Laplace transform which $u$ is a parameter in this new space, therefore we have
\begin{eqnarray}
u \mathsf{F}(s,u)-f_{0}(s)=\Phi_{f}^{LO}(s)\mathsf{f}(s,u)+\Theta_{f}^{LO}(s)\mathsf{G}(s,u)+\nonumber
\end{eqnarray}
 \begin{equation}
\Upsilon_{f}^{LO}(s)\left(a_{10}\mathsf{F}(s,u)+a_{11}\mathsf{F}(s,u-b_{11})\right)+\Omega_{f}^{LO}(s)\left(a_{10}\mathsf{M}(s,u)+a_{11}\mathsf{M}(s,u-b_{11})\right),
 \end{equation}
 \begin{equation}
 u\mathsf{G}(s,u)-g_{0}(s)=\Phi_{g}^{LO}(s)\mathsf{G}(s,u)+\Theta_{g}^{LO}(s)\mathsf{F}(s,u),
 \end{equation}
 \begin{equation}
u\mathsf{M}(s,u)-m_{0}(s)=\Omega_{m}^{LO}(s)\left(a_{10}\mathsf{M}(s,u)+a_{11}\mathsf{M}(s,u-b_{11})\right)+\Upsilon_{m}^{LO}(s)\left(a_{10}\mathsf{F}(s,u)+a_{11}\mathsf{F}(s,u-b_{11})\right).
 \end{equation}
where the Laplace transforms are as
  \begin{equation}
\mathsf{H}(s,u)=\mathcal{L}\left[h(s,\tau);u\right],\ \
\mathsf{H}(s,u-c)=\mathcal{L}\left[h(s,\tau)\exp(c\tau);u\right], \ \ \mathsf{H}(s,u+c)=\mathcal{L}\left[h(s,\tau)\exp(-c\tau);u\right].
 \end{equation}
The above equations can be easily solved by setting  $a_{11}=0$ in Eq. (23) which called as the first approximation of function $\frac{\alpha_{e}(Q^{2})}{\alpha_{s}^{LO}(Q^{2})}$. This approximation, lead us to,
 \begin{equation}
\left(u-\Phi_{f}^{LO}(s)-\Upsilon_{f}^{LO}(s)a_{10}\right) \mathsf{F}_{1}(s,u)-\Theta_{f}^{LO}(s)\mathsf{G}_{1}(s,u)-\Omega_{f}^{LO}(s)a_{10}\mathsf{M}_{1}(s,u)=f_{0}(s),
 \end{equation}
 \begin{equation}
\left(u-\Phi_{g}^{LO}(s)\right)\mathsf{G}_{1}(s,u)+\Theta_{g}^{LO}(s)\mathsf{F}_{1}(s,u)=g_{0}(s),
 \end{equation}
 \begin{equation}
\left(u-\Omega_{m}^{LO}(s)a_{10}\right)\mathsf{M}_{1}(s,u)-\Upsilon_{m}^{LO}a_{10}\mathsf{F}_{1}(s,u)=m_{0}(s).
 \end{equation}
 One can solve these equations and obtain $ \mathsf{F}_{1}$, $ \mathsf{G}_{1}$ and $\mathsf{M}_{1}$ distributions.
  \begin{equation}
\mathsf{F}_{1}(s,u)=A_{1}f_{0}(s)+A_{2}g_{0}(s)+A_{3}m_{0}(s),
\end{equation}
 \begin{equation}
\mathsf{G}_{1}(s,u)=B_{1}f_{0}(s)+B_{2}g_{0}(s)+B_{3}m_{0}(s),
\end{equation}
 \begin{equation}
\mathsf{M}_{1}(s,u)=C_{1}f_{0}(s)+C_{2}g_{0}(s)+C_{3}m_{0}(s),
\end{equation}
The coefficients $A_{1}...C_{3}$ are given in Appendix. Using the inverse Laplace transform, one can transfer the obtained equations from $u$ space to $\tau$ space, the results clearly based on the input the singlet, gluon and photon distribution functions at $Q_{0}^{2}$. The inverse transforms of $\mathsf{F}_{1}(s,u)$, $\mathsf{G}_{1}(s,u)$, and $\mathsf{M}_{1}(s,u)$ denoted by $f_{1}(s, \tau)$, $g_{1}(s, \tau)$, and $m_{1}(s, \tau)$ and therefore these functions are well defined and simple to calculate which can be expressed as
\begin{equation}
f_{1}(s, \tau)=l_{f1}(s,\tau)f_{0}(s)+l_{f2}(s,\tau)g_{0}(s)+l_{f3}(s,\tau)m_{0}(s),
\end{equation}
\begin{equation}
g_{1}(s, \tau)=l_{g1}(s,\tau)f_{0}(s)+l_{g2}(s,\tau)g_{0}(s)+l_{g3}(s,\tau)m_{0}(s),
\end{equation}
\begin{equation}
m_{1}(s, \tau)=l_{m1}(s,\tau)f_{0}(s)+l_{m2}(s,\tau)g_{0}(s)+l_{m3}(s,\tau)m_{0}(s),
\end{equation}
where, the inverse Laplace transform of cofficients $A_{i}$, $B_{i}$ and $C_{i}$ in above equations from $u$ space to $\tau$ space defined  as
$$
l_{fi}(s,\tau)=\mathcal{L}^{-1}\left[A_{i}(s,u);\tau\right]=\frac{1}{2\pi i}\int_{c-i\infty}^{c+i\infty}A_{i}(s,u)\exp(u\tau)du,
$$
$$
l_{gi}(s,\tau)=\mathcal{L}^{-1}\left[B_{i}(s,u);\tau\right],
$$
\begin{equation}
 \ \ \ \ \ \ \ \ \ \ \ \ \ \ \ \ \ \ \ \ \ \ l_{mi}(s,\tau)=\mathcal{L}^{-1}\left[C_{i}(s,u);\tau\right].    \ \ \ \ \ \ \ \ \ i=1,2,3
\end{equation}
where $c$ is a real constant such that the integration contour lies to the right of all singularities of $A_{i}(s,u)$. We re-solve the Eqs. (24-26) to obtain the
next approximation ($a_{11}\neq 0$) $\mathsf{F}_{2}$, $\mathsf{G}_{2}$, and $\mathsf{M}_{2}$ for $\mathsf{F}$, $\mathsf{G}$, $\mathsf{M}$ and then repeat the process

\begin{equation}
\left(u-\Phi_{f}^{LO}(s)-\Upsilon_{f}^{LO}(s)a_{10}\right) \mathsf{F}_{2}(s,u)-\Theta_{f}^{LO}(s)\mathsf{G}_{2}(s,u)-\Omega_{f}^{LO}(s)a_{10}\mathsf{M}_{2}(s,u)=f_{0}{'}(s),
 \end{equation}
 \begin{equation}
\left(u-\Phi_{g}^{LO}(s)\right)\mathsf{G}_{2}(s,u)+\Theta_{g}^{LO}(s)\mathsf{F}_{2}(s,u)=g_{0}(s),
 \end{equation}
 \begin{equation}
\left(u-\Omega_{m}^{LO}(s)a_{10}\right)\mathsf{M}_{2}(s,u)-\Upsilon_{m}^{LO}a_{10}\mathsf{F}_{2}(s,u)=m_{0}^{'}(s),
 \end{equation}
where
\begin{equation}
f_{0}{'}(s)=f_{0}(s)+a_{11}\left(\Upsilon_{f}^{LO}\mathsf{F}_{1}(s,u-b_{11})+\Omega_{f}^{LO}(s)\mathsf{M}_{1}(s,u-b_{11})\right),
\end{equation}
\begin{equation}
m_{0}{'}(s)=m_{0}(s)+a_{11}\left(\Omega_{m}^{LO}(s)\mathsf{M}_{1}(s,u-b_{11})+\Upsilon_{m}^{LO}(s)\mathsf{F}_{1}(s,u-b_{11})\right),
\end{equation}
similar to the previous approximation ($a_{11}=0$), one can solve these equations and obtain the $\mathsf{F}_{2}$, $\mathsf{G}$ and $\mathsf{M}$ as follows
\begin{equation}
\mathsf{F}_{2}(s,u)=A_{1}{'}f_{0}(s)+A_{2}{'}g_{0}(s)+A_{3}{'}m_{0}(s),
\end{equation}
 \begin{equation}
\mathsf{G}_{2}(s,u)=B_{1}{'}f_{0}(s)+B_{2}{'}g_{0}(s)+B_{3}{'}m_{0}(s),
\end{equation}
 \begin{equation}
\mathsf{M}_{2}(s,u)=C_{1}{'}f_{0}(s)+C_{2}{'}g_{0}(s)+C_{3}{'}m_{0}(s),
\end{equation}
where $A_{1}{'}...C_{3}{'}$ are given in Appendix
\subsection*{III. Master formula for NLO corrections }
The QCD$\otimes$QED  coupled DGLAP equations at NLO approximation in QCD for the evolution of singelt, gluon and photon distribution function, which are combination of the QCD coupled DGLAP equations at NLO and the QED  coupled DGLAP equations at LO approximation, using the convolution symbol $\otimes$ can be written as,

$$
\frac{\partial F^{s}(x,Q^{2})}{\partial \ln Q^{2}}=\frac{\alpha_{s}^{NLO}(Q^{2})}{2\pi}\Bigg[\left(P_{qq}^{LO}(x)+\frac{\alpha_{s}^{NLO}(Q^{2})}{2\pi}P_{qq}^{NLO}(x)\right)\otimes F^{s}(x,Q^{2})+2n_{f}\bigg(P_{qg}^{LO}(x)
$$
\begin{equation}
+\frac{\alpha_{s}^{NLO}(Q^{2})}{2\pi}P_{qg}^{NLO}(x)\bigg) G(x,Q^{2})\Bigg]+\frac{\alpha_{e}(Q^{2})}{2\pi}\bigg[ \frac{5}{18} \tilde{P}_{qq}^{LO}(x)\otimes F^{s} (x,Q^{2})+2A P_{q\gamma}^{LO}(x)\otimes M(x,Q^{2})\bigg],
\end{equation}

$$
\frac{\partial G(x,Q^{2})}{\partial \ln Q^{2}}=\frac{\alpha_{s}^{NLO}(Q^{2})}{2\pi}\bigg[\left(P_{gq}^{0}(x)+\frac{\alpha_{s}^{NLO}(Q^{2})}{2\pi}P_{gq}^{NLO}(x)\right)\otimes F^{s}(x,Q^{2})+\bigg(P_{gg}^{LO}(x)
$$
\begin{equation}
+\frac{\alpha_{s}^{NLO}(Q^{2})}{2\pi}P_{gg}^{NLO}(x)\bigg) G(x,Q^{2})\bigg],
\end{equation}
\begin{equation}
\frac{\partial M(x,Q^{2})}{\partial \ln Q^{2}}=\frac{\alpha_{e}(Q^{2})}{2\pi}\bigg[\frac{5}{18} P_{\gamma q}^{LO}(x)\otimes F^{s}(x,Q^{2})+P_{\gamma \gamma}^{LO}\otimes M(x,Q^{2})\bigg].
\end{equation}
To calculate above equations at NLO approximation, we consider the following expression for $\frac{\alpha_{s}^{NLO}(Q^{2})}{4\pi}$ and $\frac{\alpha_{e}(Q^{2})}{\alpha_{s}^{NLO}(Q^{2})}$ as,
\begin{equation}
\frac{\alpha_{e}(Q^{2})}{\alpha_{s}^{NLO}(Q^{2})}\approx a_{20}+a_{21}\exp(b_{21}\tau),\\\\\ \frac{\alpha_{s}^{NLO}(Q^{2})}{4\pi}\approx a_{30}+a_{31}\exp(-b_{31}\tau)
\end{equation}
as shown in Table (1), this expression includes an excellent accurate. The running coupling constant at NLO is defined as,
\begin{eqnarray}
 \alpha_{s}^{NLO}\left(Q^{2}
\right)=\frac{4\pi}{\beta_{0}\ln\left(\frac{Q^{2}}{\Lambda^{2}}\right)}\left(1-\frac{\beta_{1}}{\beta_{0}}\frac{\ln\left(\ln\left(\frac{Q^{2}}{\Lambda^{2}}\right)\right)}{\ln\left(\frac{Q^{2}}{\Lambda}\right)}\right),
\end{eqnarray}
where $\beta_{1}=102-\frac{38}{2}n_{f}$ is NLO correction to the QCD $\beta$-
function. to solve Eqs. (47-49), we again need double Laplace  transformation from $x$ to $s$ and $\tau$ to $u$ space. Therefore the solution of the differential-integral equations in Eqs. (47-49) can be converted to,
\begin{eqnarray}
u \mathsf{F}^{NLO}(s,u)-f^{NLO}_{0}(s)=\Phi_{f}^{LO}(s)\mathsf{F}^{NLO}(s,u)+\Theta_{f}^{LO}(s)\mathsf{G}^{NLO}(s,u)+\nonumber
\end{eqnarray}
\begin{eqnarray}
\Phi_{f}^{NLO}(s)\left(a_{30}\mathsf{F}^{NLO}(s,u)+a_{31}\mathsf{F}^{NLO}(s,u+b_{31})\right)+\Theta_{f}^{NLO}(s)\left(a_{30}\mathsf{G}(s,u)+a_{31}\mathsf{G}^{NLO}(s,u+b_{31})\right)+\nonumber
\end{eqnarray}
 \begin{equation}
\Upsilon_{f}^{LO}(s)\left(a_{20}\mathsf{F}^{NLO}(s,u)+a_{21}\mathsf{F}^{NLO}(s,u-b_{21})\right)+\Omega_{f}^{LO}(s)\left(a_{20}\mathsf{M}^{NLO}(s,u)+a_{21}\mathsf{M}^{NLO}(s,u-b_{21})\right),
 \end{equation}
$$
 u\mathsf{G}^{NLO}(s,u)-g_{0}(s)=\Phi_{g}^{LO}(s)\mathsf{G}^{NLO}(s,u)+\Theta_{g}^{LO}(s)\mathsf{F}^{NLO}(s,u)+\Theta_{g}^{NLO}(s)\big(a_{30}\mathsf{F}^{NLO}(s,u)
$$
 \begin{equation}
+a_{31}\mathsf{F}^{NLO}(s,u+b_{31})\big)+ \Phi_{g}^{NLO}(s)\left(a_{30}\mathsf{G}^{NLO}(s,u)+a_{31}\mathsf{G}^{NLO}(s,u+b_{31})\right),
 \end{equation}
 $$
u\mathsf{M}^{NLO}(s,u)-m_{0}(s)=\Omega_{m}^{LO}(s)\left(a_{20}\mathsf{M}^{NLO}(s,u)+a_{21}\mathsf{M}^{NLO}(s,u-b_{21})\right)
$$
\begin{equation}
+\Upsilon_{m}^{LO}(s)\left(a_{20}\mathsf{F}(s,u)+a_{21}\mathsf{F}(s,u-b_{21})\right).
 \end{equation}
The NLO splitting functions $\Phi_{f}^{NLO}$, $\Theta_{f}^{NLO}$, $\Phi_{g}^{NLO}$ $\Theta_{g}^{NLO}$ can be easily obtained in $s$ space using the NLO results drived in Refs. \cite{27,28,29} as Ref. \cite{30}. By setting $a_{21}=0$ and $a_{31}=0$ in Eqs. (49), the above equations can be obtained  at first approximation as,

$$
\left(u-\Phi_{f}^{LO}(s)-a_{30}\Phi_{f}^{NLO}(s)-\Upsilon_{f}^{LO}(s)a_{20}\right) \mathsf{F}_{1}^{NLO}(s,u)-\left(\Theta_{f}^{LO}(s)+a_{30}\Theta_{f}^{NLO}(s)\right)\mathsf{G}_{1}^{NLO}(s,u)
$$
\begin{equation}
-\Omega_{f}^{LO}(s)a_{20}\mathsf{M}_{1}^{NLO}(s,u)=f_{0}^{NLO}(s),
 \end{equation}
 \begin{equation}
\left(u-\Phi_{g}^{LO}(s)-a_{30}\Phi_{g}^{NLO}(s)\right)\mathsf{G}_{1}^{NLO}(s,u)-\left(\Theta_{g}^{LO}(s)+a_{30}\Theta_{g}^{NLO}(s)\right)\mathsf{F}_{1}^{NLO}(s,u)=g_{0}^{NLO}(s),
 \end{equation}
 \begin{equation}
\left(u-\Omega_{m}^{LO}(s)a_{20}\right)\mathsf{M}_{1}^{NLO}(s,u)-\Upsilon_{m}^{LO}a_{20}\mathsf{F}_{1}^{NLO}(s,u)=m_{0}(s).
 \end{equation}
 One can solve these equations and obtain $ \mathsf{F}_{1}^{NLO}$, $ \mathsf{G}_{1}^{NLO}$ and $\mathsf{M}_{1}^{NLO}$ distributions, the results depend on the input singlet, gluon and photon distribution functions at initial scale,
  \begin{equation}
\mathsf{F}_{1}^{NLO}(s,u)=A_{1}^{NLO}f_{0}^{NLO}(s)+A_{2}^{NLO}g_{0}^{NLO}(s)+A_{3}^{NLO}m_{0}(s),
\end{equation}
 \begin{equation}
\mathsf{G}_{1}^{NLO}(s,u)=B_{1}^{NLO}f_{0}^{NLO}(s)+B_{2}^{NLO}g_{0}^{NLO}(s)+B_{3}^{NLO}m_{0}(s),
\end{equation}
 \begin{equation}
\mathsf{M}_{1}^{NLO}(s,u)=C_{1}^{NLO}f_{0}^{NLO}(s)+C_{2}^{NLO}g_{0}^{NLO}(s)+C_{3}^{NLO}m_{0}(s),
\end{equation}
 to obtain the next approximation ($a_{21}\neq 0,\ a_{31}\neq 0$) $\mathsf{F}_{2}^{NLO}$, $\mathsf{G}_{2}^{NLO}$, and $\mathsf{M}_{2}^{NLO}$ for $\mathsf{F}$, $\mathsf{G}$, $\mathsf{M}$, we replace $f_{0}{'}^{NLO}$, $g_{0}{'}^{NLO}$ and $m_{0}{'}^{NLO}$ with $f_{0}^{NLO}$,$g_{0}^{NLO}$ and $m_{0}$ respectively in Eqs. (54-56). The method of calculating next approximation is as first approximation.  $f_{0}{'}^{NLO}$, $g_{0}{'}^{NLO}$ and $m_{0}{'}^{NLO}$  are as follows
$$
f_{0}{'}^{NLO}(s,u)=f_{0}^{NLO}(s)+a_{21}\left(\Upsilon_{f}^{LO}(s)\mathsf{F}_{1}^{NLO}(s,u-b_{21})+\Omega_{f}^{LO}(s)\mathsf{M}_{1}^{NLO}(s,u-b_{21})\right)
$$
\begin{equation}
+a_{31}\left(\Phi_{f}^{NLO}\mathsf{F}_{1}^{NLO}(s,u+b_{31})+\Theta_{f}^{NLO}(s)\mathsf{G}_{1}^{NLO}(s,u+b_{31})\right)
\end{equation}
\begin{equation}
g_{0}{'}^{NLO}(s,u)=g_{0}^{NLO}(s)+a_{31}\left(\Theta_{g}^{NLO}\mathsf{F}_{1}^{NLO}(s,u+b_{31})+\Phi_{f}^{NLO}(s)\mathsf{G}_{1}^{NLO}(s,u+b_{31})\right)
\end{equation}
\begin{equation}
m_{0}{'}^{NLO}(s,u)=m_{0}(s)+a_{21}\left(\Omega_{m}^{LO}(s)\mathsf{M}_{1}(s,u-b_{21})+\Upsilon_{m}^{LO}(s)\mathsf{F}_{1}(s,u-b_{21})\right),
\end{equation}
  one can obtain the $\mathsf{F}_{2}^{NLO}$, $\mathsf{G}_{2}^{NLO}$ and $\mathsf{M}_{2}^{NLO}$ from Eqs. (57-62) as follows
\begin{equation}
\mathsf{F}_{2}^{NLO}(s,u)=A_{1}{'}^{NLO}f_{0}^{NLO}(s)+A_{2}{'}^{NLO}g_{0}^{NLO}(s)+A_{3}^{NLO}{'}m_{0}(s),
\end{equation}
 \begin{equation}
\mathsf{G}_{2}^{NLO}(s,u)=B_{1}{'}^{NLO}f_{0}^{NLO}(s)+B_{2}{'}^{NLO}g_{0}^{NLO}(s)+B_{3}^{NLO}{'}m_{0}(s),
\end{equation}
 \begin{equation}
\mathsf{M}_{2}^{NLO}(s,u)=C_{1}{'}^{NLO}f_{0}^{NLO}(s)+C_{2}{'}^{NLO}g_{0}^{NLO}(s)+C_{3}{'}^{NLO}m_{0}(s),
\end{equation}
where $A_{1}{'}^{NLO}...C_{3}{'}^{NLO}$ are given in Appendix. Using iterative solution of Eqs. (43-45) and Eqs. (63-65) and  the inverse Laplace transform technique for back from $u$ to $\tau$ space, the following expressions for the singlet, gluon and photon distributions can be obtained as
 \begin{equation}
f^{n}(s, \tau)=k_{f1}^{n}(s,\tau)f_{0}^{n}(s)+k_{f2}^{n}(s,\tau)g_{0}^{n}(s)+k_{f3}^{n}(s,\tau)m_{0}(s),\ \ \\\\
\end{equation}
\begin{equation}
g^{n}(s, \tau)=k_{g1}^{n}(s,\tau)f_{0}^{n}(s)+k_{g2}^{n}(s,\tau)g_{0}^{n}(s)+k_{g3}^{n}(s,\tau)m_{0}(s),\ \ \\\\
\end{equation}
\begin{equation}
m^{n}(s, \tau)=k_{m1}^{n}(s,\tau)f_{0}^{n}(s)+k_{m2}^{n}(s,\tau)g_{0}^{n}(s)+k_{m3}^{n}(s,\tau)m_{0}(s), \ \ \\\\
\end{equation}
$$
n=LO or NLO
$$
where the coefficients at  $\tau$ space are,
$$
k_{fi}^{n}(s, \tau)=\mathcal{L}^{-1}\left[A_{i}{'}^{n}(s,u);\tau\right],\ \ \ \quad \quad\quad \quad \ \ \ \quad \quad
$$
$$
k_{gi}^{n}(s, \tau)=\mathcal{L}^{-1}\left[B_{i}{'}^{n}(s,u);\tau\right],\ \ \ \quad \quad\quad \quad \ \ \ \quad \quad
$$
\begin{equation}
 \ \ \ \ \ \ \ \ \ \ \ \ \ \ \ \ \ \ \ \ \ \ \ \ k_{mi}^{n}(s, \tau)=\mathcal{L}^{-1}\left[C_{i}{'}^{n}(s,u);\tau\right], \quad  i=1,2,3 \ \ \ and \ \  n=LO or NLO
\end{equation}
one can invert $k_{ji}^{n}(s)$ ($j=f, g, m$) from $s$ space to $\upsilon$ space
using algorithms of Ref. \cite{24}, there is showed that the real
basis of the method is in assuring that the inverse transforms can
be calculated exactly to high orders (as defined there), even for
function which diverge for $s\rightarrow 0$. We define their
inverse Laplace  as,
 $$
K_{fi}^{n}(\upsilon,\tau)=\mathcal{L}^{-1}\left[k_{fi}^{n}(s,\tau);\upsilon\right],\ \ \ \quad \quad\quad \quad \ \ \ \quad \quad
$$
$$
K_{gi}^{n}(\upsilon,\tau)=\mathcal{L}^{-1}\left[k_{gi}^{n}(s,\tau);\upsilon\right],\ \ \ \quad \quad\quad \quad \ \ \ \quad \quad
$$
\begin{equation}
\ \ \ \ \ \ \ \ \ \ \ \ \ \ \ \ \ \ \ \ \ \ K_{mi}^{n}(\upsilon,\tau)=\mathcal{L}^{-1}\left[k_{mi}^{n}(s,\tau);\upsilon\right], \quad  i=1,2,3 \ \ \ and \ \  n=LO or NLO
\end{equation}
 so, we can write the solutions in $(\upsilon, \tau)$ space as the convolutions,
$$
\hat{F}^{s\ n}(v,\tau)=\int_{0}^{\upsilon}K_{f1}^{n}\left(v-w,\tau\right)\hat{F}_{0}^{s\ n}\left(w\right)dw+\int_{0}^{\upsilon}K_{f2}^{n}\left(v-w,\tau\right)\hat{G}^{n}_{0}\left(w\right)dw+
$$
\begin{equation}
\int_{0}^{\upsilon}K_{f3}^{n}\left(v-w,\tau\right)\hat{M}^{n}_{0}\left(w\right)dw,
\end{equation}
$$
\hat{G}^{n}(v,\tau)=\int_{0}^{\upsilon}K_{g1}^{n}\left(v-w,\tau\right)\hat{F}_{0}^{s\ n}\left(w\right)dw+\int_{0}^{\upsilon}K_{g2}^{n}\left(v-w,\tau\right)\hat{G}_{0}^{n}\left(w\right)dw+
$$
\begin{equation}
\int_{0}^{\upsilon}K_{g3}^{n}\left(v-w,\tau\right)\hat{M}_{0}^{n}\left(w\right)dw,
\end{equation}
$$
\hat{M}^{n}(v,\tau)=\int_{0}^{\upsilon}K_{l1}^{n}\left(v-w,\tau\right)\hat{F}_{0}^{s\ n}\left(w\right)dw+\int_{0}^{\upsilon}K_{l2}^{n}\left(v-w,\tau\right)\hat{G}_{0}^{n}\left(w\right)dw+
$$
\begin{equation}
\int_{0}^{\upsilon}K_{f3}^{n}\left(v-w,\tau\right)\hat{M}^{n}_{0}\left(w\right)dw. \ \ \ \ \  n=LO or NLO
\end{equation}
Finally, recalling the $\upsilon \equiv\ ln(1/x)$, one can convert the above solutions back into the usual space, Bjorken-$x$
and virtuality $Q^{2}$. The $Q^{2}$ dependence of the solutions are performed by the $\tau$ variable. Consequently, we can obtain the singlet, gluon and photon distribution as $F^{s}(x,Q^{2})$,  $G(x,Q^{2})$ and $M(x,Q^{2})$.  Eq. (73) indicates that photon contribution in $Q^{2}{‘}$s larger than initial scale depends on gluon contribution at initial condition.
\subsection*{IV. Results and Conclusion}
In this section, we will present the results that have been
obtained for the proton structure function and singlet, gluon and
photon distribution functions by using the double Laplace
transform technique to find an analytical solution for the
QCD$\otimes$ QED  DGLAP evolution equations at LO and NLO QCD. Also we give an
example to this approach and compare the proton structure function at NLO approximation
with APFEL results, this function calculated with a suitable
approximation from the singlet distribution function. In addition,
the singlet and gluon distribution functions, obtained from Eqs.
(71, 72)  starting from the MSTW initial conditions at
$Q_{0}^{2}=1~GeV^{2}$ \cite{9}, are compared with NLO MSTW
distributions and the photon distribution function with MRST
\cite{11}. Initial condition for the photon distribution function
is gained at $Q_{0}^{2}=1~ GeV^{2}$ by \cite{11}. In this method,
it is written relations $\frac{\alpha_{e}(Q^{2})}{\alpha_{s}(Q^{2})}$, $ \frac{\alpha_{e}(Q^{2})}{\alpha_{s}^{NLO}(Q^{2})}$ and $ \frac{\alpha_{s}^{NLO}(Q^{2})} {4\pi}$ as summation a exponential term (which is a function of $\tau$) and a constant number, in Table (1) we showed $a_{10}$ ... $b_{31}$ from $ Q^{2}=3 ~GeV^{2}$ to $8000~ GeV^{2}$ with a
good fitting. In Figs. (2-4), it is compared the results of these functions with Eqs. (23,49). As these figures show, the used expansions are appropriate for the solution of DGLAP equations. Fig. (5) shows the results of the proton structure
function at  NLO approximation which is compared with MSTW and APFEL results  at
$Q^{2}=120$, $1000$ and $8000$  $GeV^{2}$ scales,  in Figs. (6) we compared
the our results at NLO QCD with HERA data, calculations show that NLO corrections are very close to the other results and data. Although the contribution of photon in the
proton structure function is very small, but it is larger than
contribution of $b$ quark at low $Q^{2}$ and large $x$ as shown in Ref. \cite{11}. Figs. (7, 8)
show the results singlet and gluon distribution functions at leading order and next-to-leading order obtained
from Eqs. (71, 72) in $x$-space at LO and NLO in QCD for $Q^{2}$ values of (20,
100, $M_{z}^{2}~GeV^{2}$) and compare these results with MSTW. The
singlet and gluon distributions are very similar to the NLO standard
MSTW distributions, where $F^{s}_{0}$ and $G_{0}$ are the MSTW
values at initial scale. Solid lines are the our results at LO (up) and NLO (down), and dash
lines are MSTW values. In Fig. (9) is plotted the photon
distribution function obtained from Eq. (73) in $x$-space in LO and NLO in QCD at
different energy scales and it compared with  MRST at
$Q^{2}=20~GeV^{2}$. This figure shows that the obtained results
from present analysis based on Laplace transform technique are in
good agreements with the ones obtained by MRST. To understand the
effect of photon distribution function on the proton structure
function, in Figs. (10,11)  this distribution function is compared
with the sea quarks distribution functions inside the proton which are presented by CETQ
\cite{14}. In Figs. (10) we showed different between the photon and
$b$ quark distribution function, as it is seen, in low energy
$Q^{2}=12, 25 ~GeV^{2}$ the photon distribution is larger than $b$
quark but by increasing energy the photon distribution both in LO and NLO corrections decreases.
It should be expressed that in range $x>0.03$ and
$Q^{2}<200~GeV^{2}$ the photon distribution function has dominant
contribution  in comparison with $b$ quark. In Figs. (11), the sea distribution functions are
larger than the photon distribution function but in high energy
and $0.1<x<1$ the contribution of photon is significant in analogy
with sea quarks.

In conclusion, we obtained three analytical decoupled differential evolution equations for the singlet and gluon and photon distribution functions from the QCD$\otimes$QED coupled DGLAP equations by using the double Laplace method at leading order and next-to-leading order. These equations are general and require only a knowledge of $F^{s}_{0}$, $G_{0}$, $M_{0}$  at the starting value $Q^{2}_{0}$ for the evolution. The most important parts of this paper was that photon distribution function at high energy scales depends on gluon distribution function at initial energy scale. Using the Laplace transform method, we obtained contribution of photon in proton at high energy and showed which this contribution is larger than b heavy quark in specific area of $x$. We showed that the photon distribution function is comparable with sea quarks distribution functions of specially at high $x$ and middle $Q^{2}$.  Also it is observed that the general solutions are in good agreement with available the experimental data and other parameterization models.
\subsection*{Acknowledgment}
G.R. Boroun thanks Prof.L. Durand for a fruitful discussion on the
inverse Laplace transforms and their accuracy. Thanks to Prof.P. Ha
for useful discussions.\\

\newpage
\begin{figure}[h]
\begin{center}
\includegraphics[width=.9\textwidth]{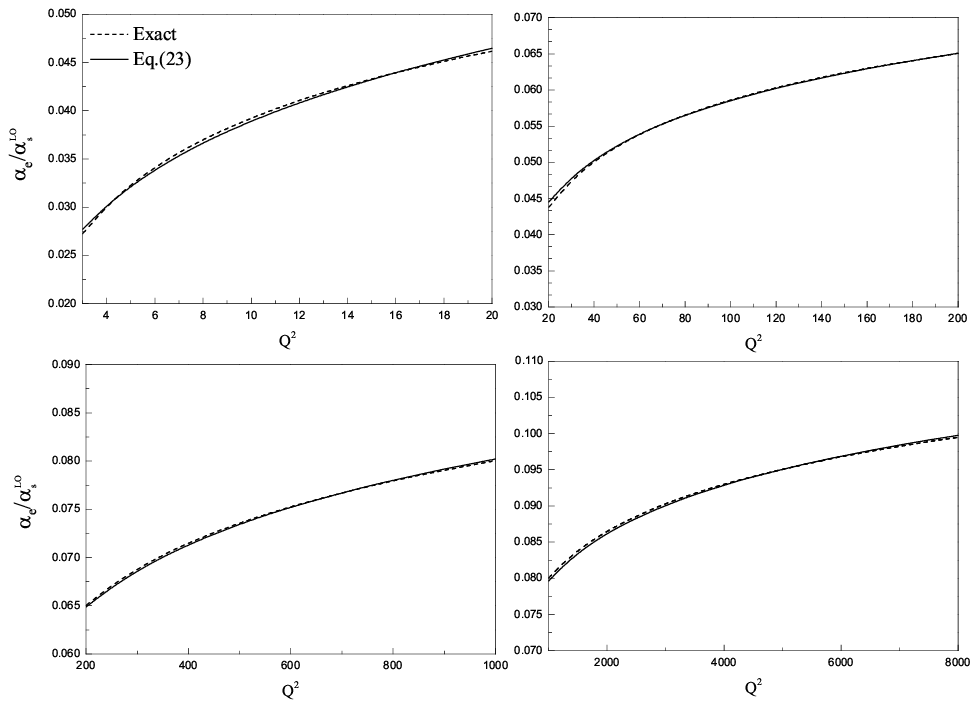}
\quad

\begingroup
    \fontsize{10pt}{12pt}\selectfont  {
\label*{
Figure 2: Solid lines show the results of Eq. (23) and dash lines show the exact values of $\frac{\alpha_{e}(Q^{2})}{\alpha_{s}^{LO}(Q^{2})}$}}.
\endgroup
\end{center}
\end{figure}

\newpage
\begin{figure}[h]
\begin{center}
\includegraphics[width=.9\textwidth]{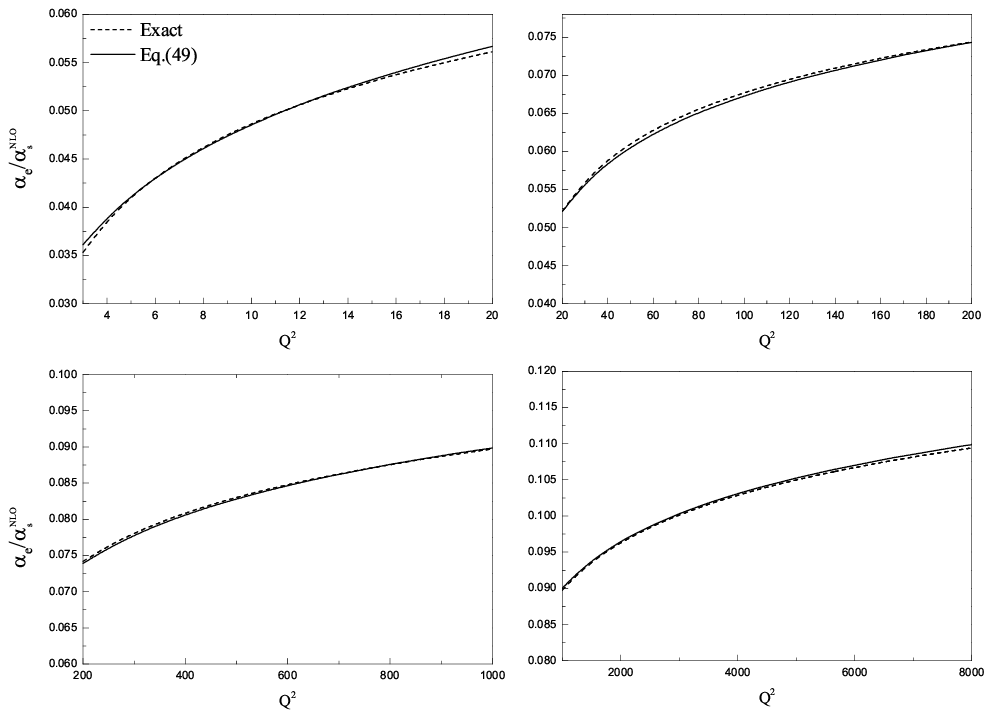}
\quad

\begingroup
    \fontsize{10pt}{12pt}\selectfont  {
\label*{
Figure 3: Solid lines show the results of Eq. (49) (left) and dash lines show the exact values of $ \frac{\alpha_{e}(Q^{2})}{\alpha_{s}^{NLO}(Q^{2})}$}}.
\endgroup
\end{center}
\end{figure}

\newpage
\begin{figure}[h]
\begin{center}
\includegraphics[width=.9\textwidth]{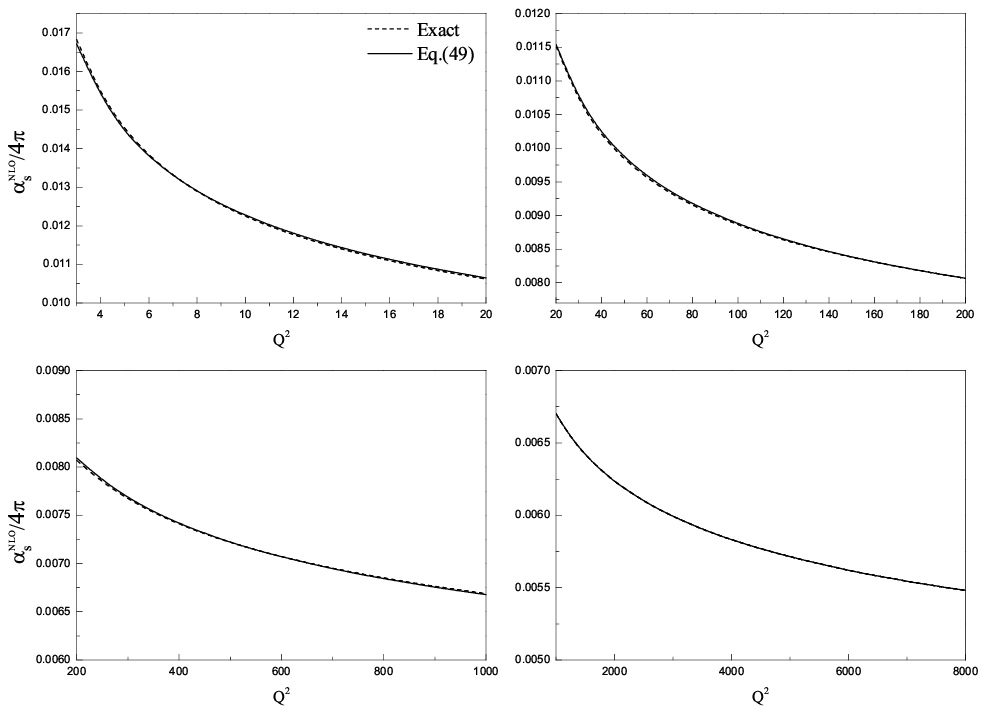}
\quad

\begingroup
    \fontsize{10pt}{12pt}\selectfont  {
\label*{
Figure 4: Solid lines show the results of Eq. (49) (right) and dash lines show the exact values of $ \frac{\alpha_{s}^{NLO}(Q^{2})} {4\pi}$  }}.
\endgroup
\end{center}
\end{figure}

\newpage
\begin{figure}[h]
\begin{center}
\includegraphics[width=.65\textwidth]{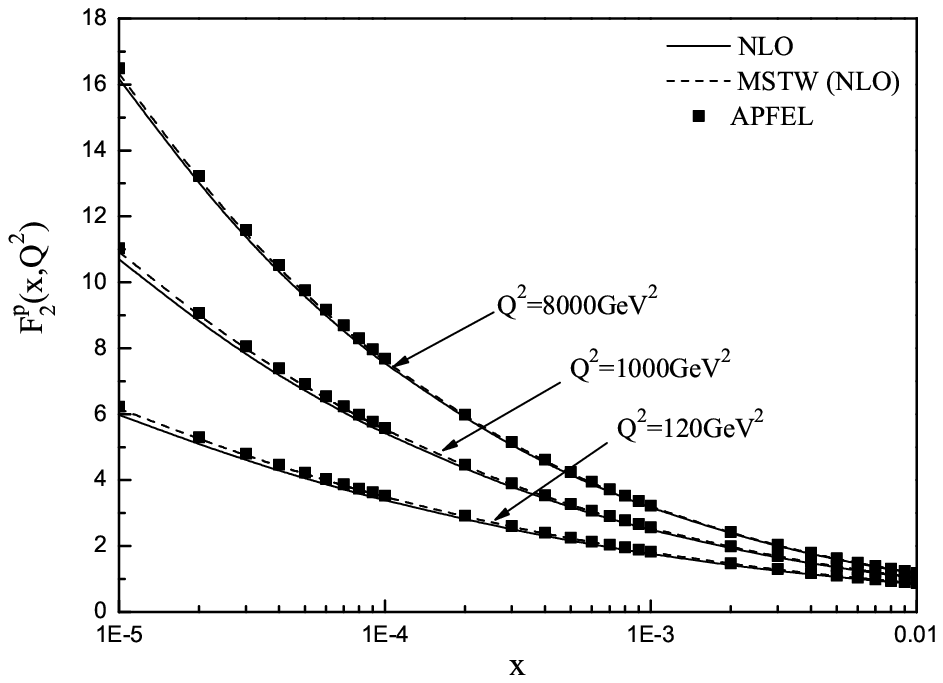}
\quad

\begingroup
    \fontsize{10pt}{12pt}\selectfont  {
\label*{
Figure 5: Solid lines show the results for the proton structure function at $Q^{2}=120$, $1000$, $8000$ $GeV^{2}$ scales in NLO, dash lines and symbols are the MSTW parameterization \cite{9} and APFEL results \cite{7} respectively}}.
\endgroup
\end{center}
\end{figure}

\begin{figure}[h]
\begin{center}
\includegraphics[width=.79\textwidth]{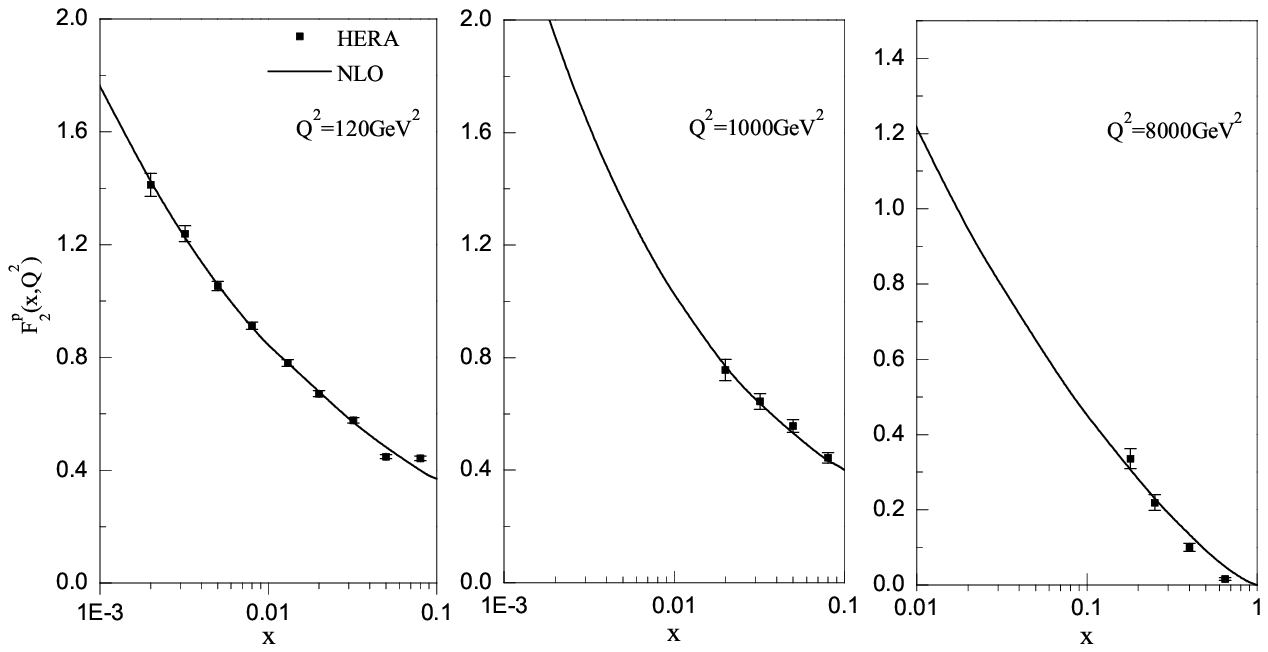}
\quad

\begingroup
    \fontsize{10pt}{12pt}\selectfont  {
\label*{
Figure 6: Solid lines show the results for the proton structure function at $Q^{2}=120$, $1000$, $8000$ $GeV^{2}$ scales in NLO and symbols are HERA data \cite{12}}}.
\endgroup
\end{center}
\end{figure}

\newpage
\begin{figure}
\begin{center}
\begin{minipage}[b]{0.7\linewidth}
\includegraphics[width=1\textwidth]{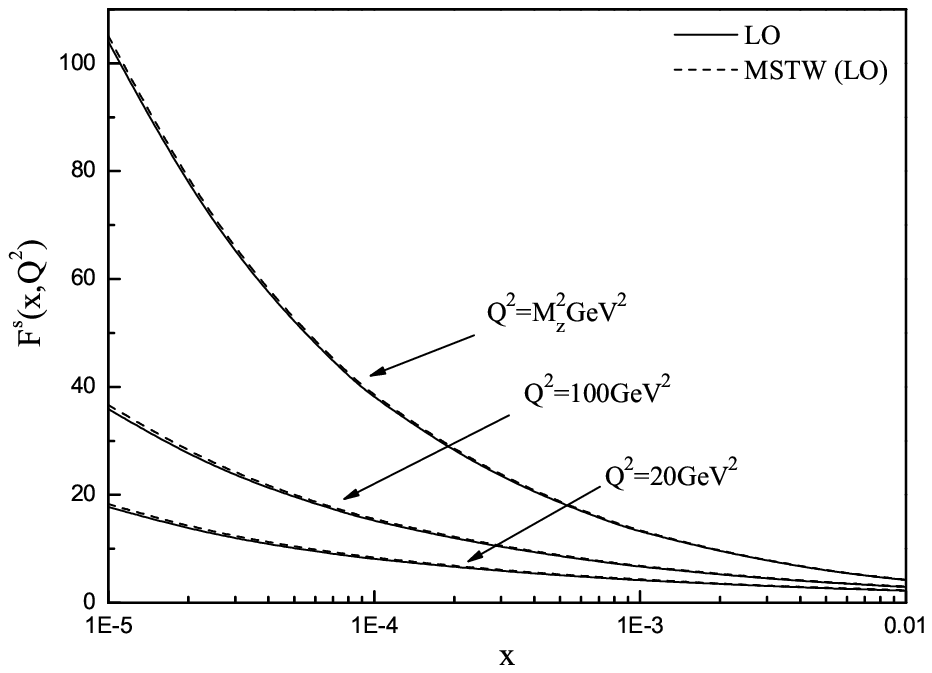}
\label{fig:minipage1}
\end{minipage}

\begin{minipage}[b]{0.7\linewidth}
\includegraphics[width=1\textwidth]{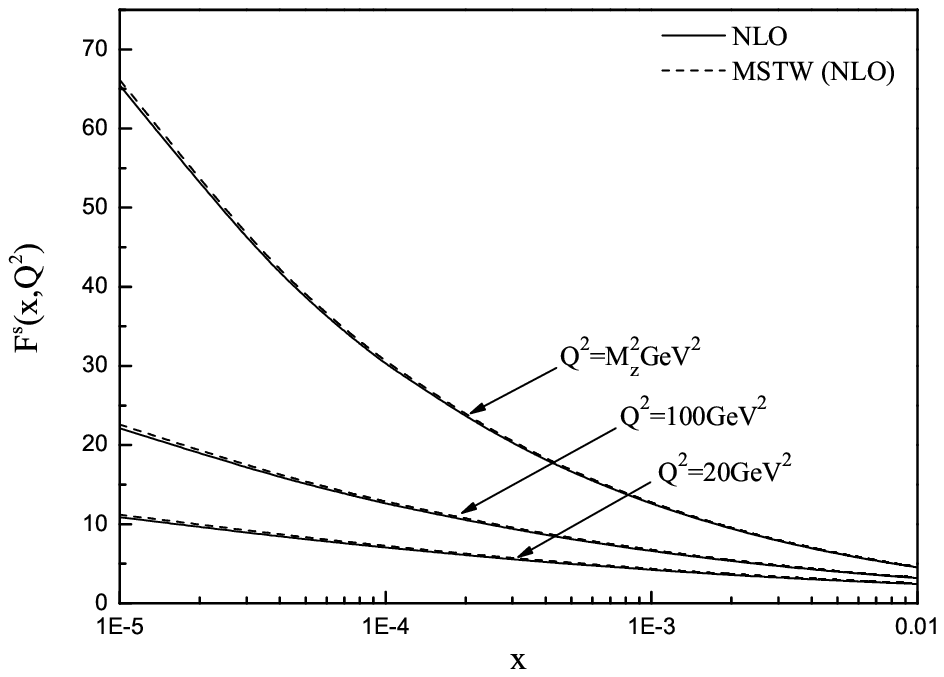}
\label{fig:minipage2}
\end{minipage}
\quad

\begingroup
    \fontsize{10pt}{12pt}\selectfont  {
\label*{
Figure 7: Solid lines show the results for singlet distribution function at $Q^{2}=20$, $100$, $M_{z}^{2}$ $GeV^{2}$ scales in LO (up) and NLO (down) and dash lines are the MSTW parameterization \cite{9}}}.
\endgroup
\end{center}
\end{figure}

\newpage
\begin{figure}
\begin{center}
\begin{minipage}[b]{0.7\linewidth}
\includegraphics[width=1\textwidth]{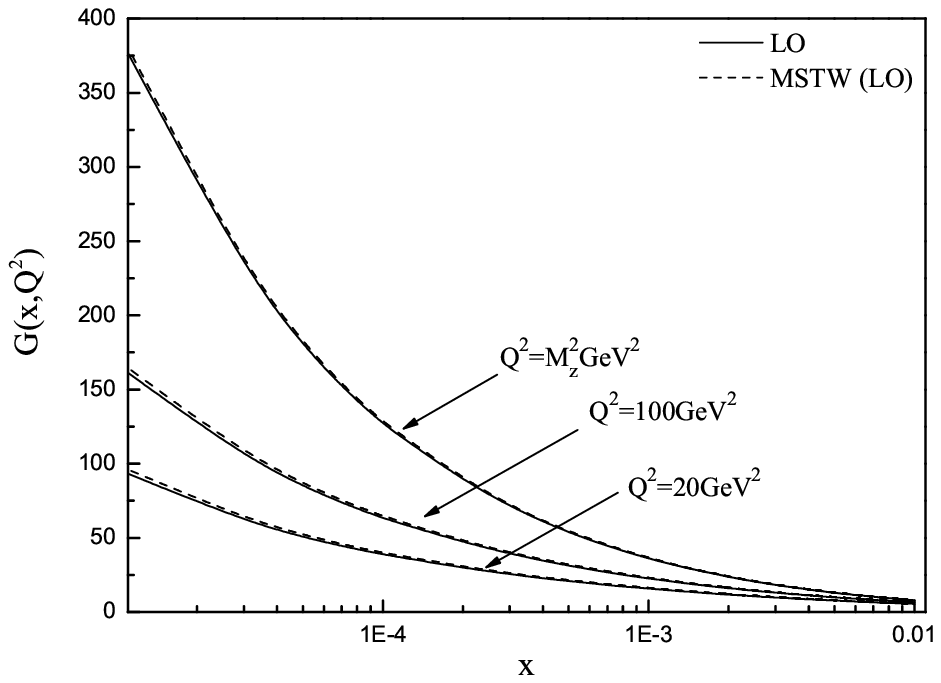}
\label{fig:minipage1}
\end{minipage}

\begin{minipage}[b]{0.7\linewidth}
\includegraphics[width=1\textwidth]{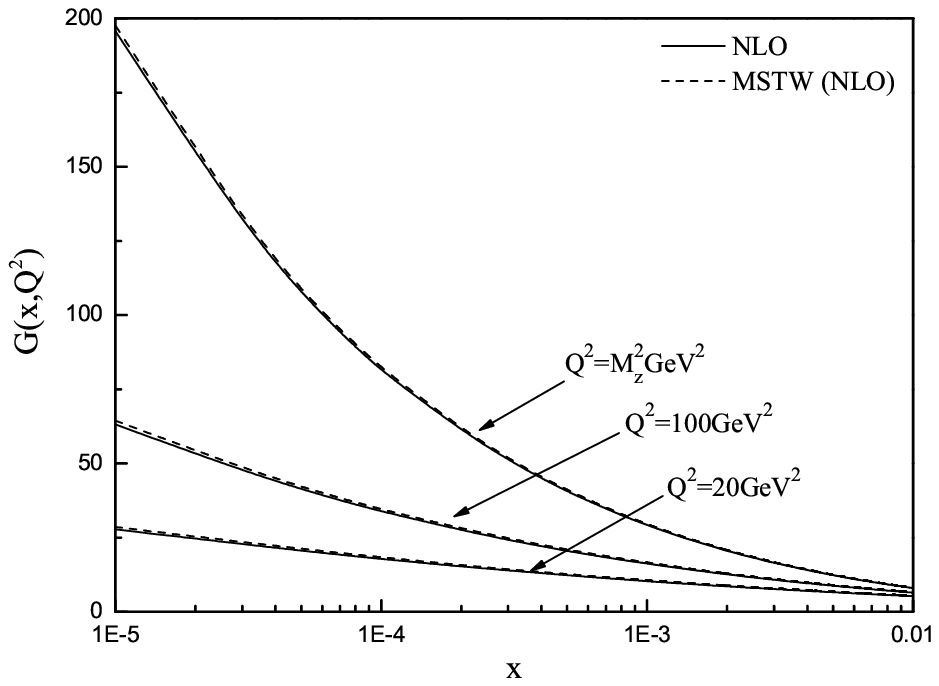}
\label{fig:minipage2}
\end{minipage}
\quad

\begingroup
    \fontsize{10pt}{12pt}\selectfont  {
\label*{
Figure 8: Solid lines show the results for gluon distribution function at $Q^{2}=20$, $100$, $M_{z}^{2}$ $GeV^{2}$ scales in LO (up) and NLO (down) and dash lines are the MSTW parameterization \cite{9}}}.
\endgroup
\end{center}
\end{figure}

\newpage
\begin{figure}
\begin{center}
\includegraphics[width=.65\textwidth]{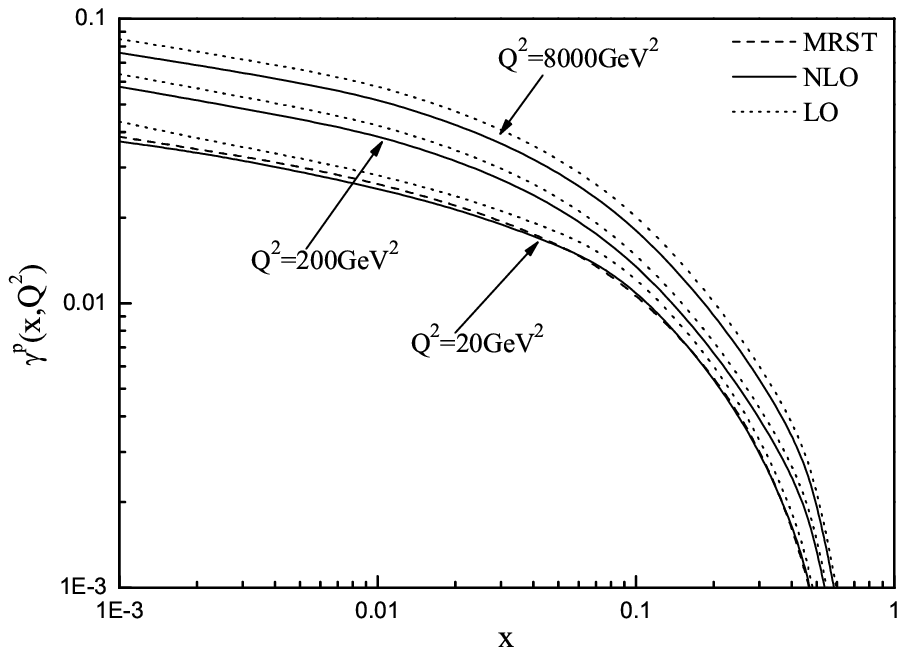}
\quad

\begingroup
    \fontsize{10pt}{12pt}\selectfont  {
\label*{
Figure 9: Solid and dot lines show the results for the photon distribution function at $Q^{2}=20, 200, 8000$ $GeV^{2}$ scales in NLO and LO, respectively and dash line is the MRST parameterization \cite{11} }}.
\endgroup
\end{center}
\end{figure}
\newpage
\begin{figure}
\begin{center}
\includegraphics[width=1.\textwidth]{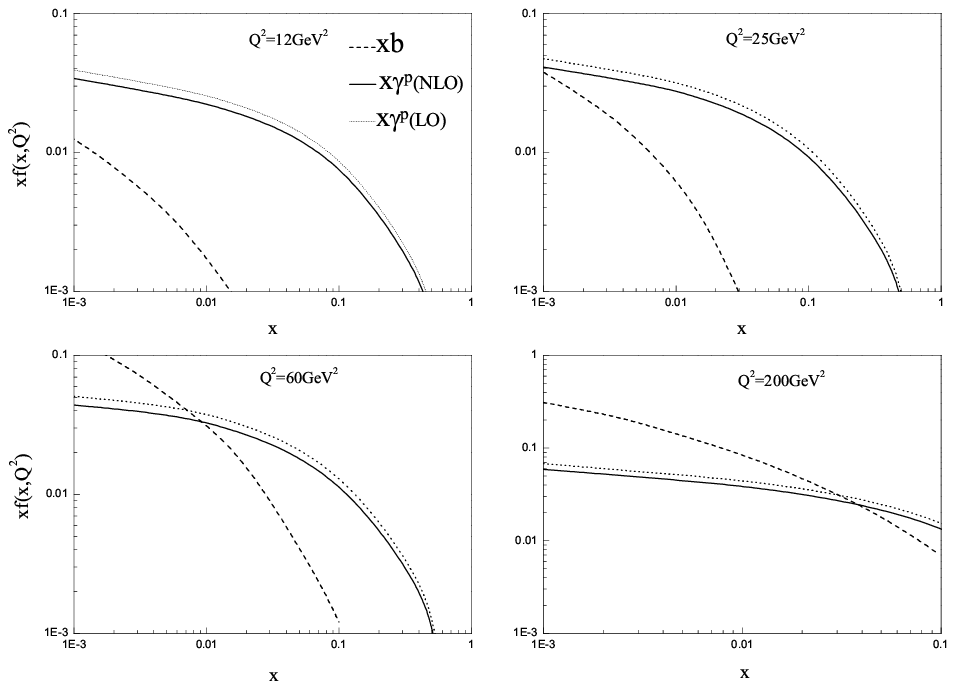}
\quad
\newpage
\begingroup
    \fontsize{10pt}{12pt}\selectfont  {
\label*{
Figure 10: Solid and dot lines show the results for the photon distribution function at $Q^{2}=12$, $25$, $60$, $200$ $GeV^{2}$ scales in NLO and LO, respectively and dash lines are $b$ quark distribution function presented by CETQ \cite{14}}}.
\endgroup
\end{center}
\end{figure}
\newpage
\begin{figure}
\begin{center}
\includegraphics[width=.8\textwidth]{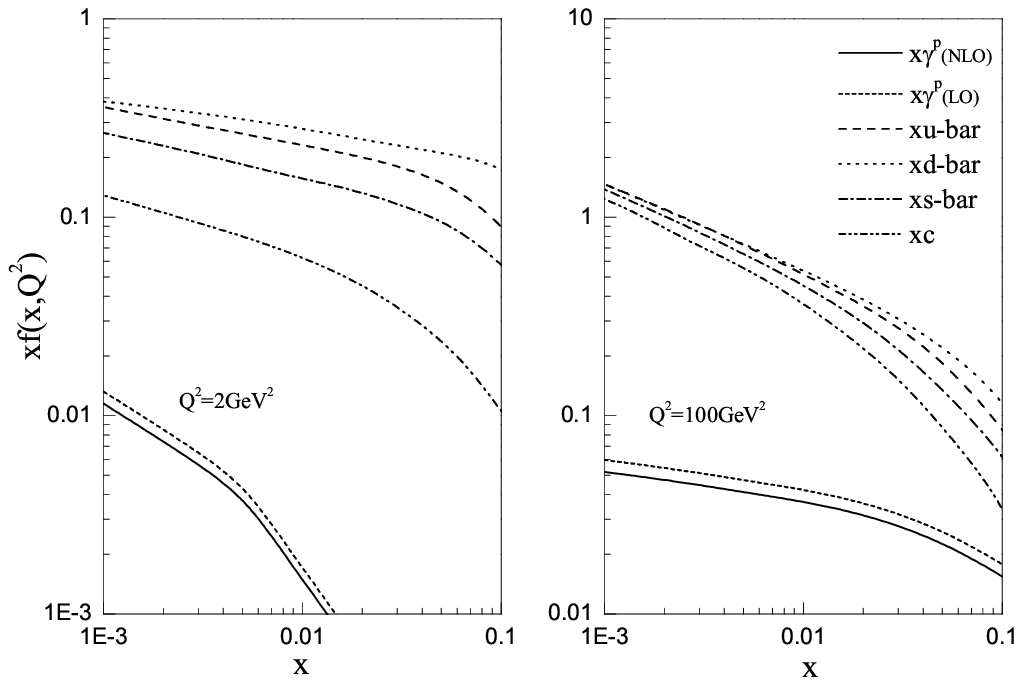}
\quad

\begingroup
    \fontsize{10pt}{12pt}\selectfont  {
\label*{ Figure 11: Solid and dot lines correspond to the results for
photon distribution function at $Q^{2}=2$, $100$ $GeV^{2}$ scales in NLO and LO, respectively and
other lines are  the distribution functions of sea quarks presented by CETQ \cite{14}}}.
\endgroup
\end{center}
\end{figure}
\newpage
\begin{table}[h]

\begingroup
\fontsize{10pt}{12pt}\selectfont{
\label*{ Table 1: Parameter values $a_{10},..., b _{31}$  at different ranges of energy}}
\endgroup
 \begin{center}
\begingroup
\fontsize{8pt}{12pt}\selectfont{
\begin{tabular} {cccccc}
\toprule[1pt]
     & &\normal{\head{$3\leq Q^{2}<20 $}}
  &\normal{\head{$20\leq Q^{2}<200 $}}\  & \normal{\head{$200\leq Q^{2}<1000 $}}& \normal{\head{$1000\leq Q^{2}<8000 GeV^{2}$}}  \\ \hline

$\frac{\alpha_{e}(Q^{2})}{\alpha_{s}^{LO}(Q^{2})}$ &$a_{10}$ \quad &$0.0088$ &  $0.0101$  &  $0.0108$&  $0.0128$\\

 & $a_{11}$  \quad  &$0.0109$ & $ 0.0134$ & $0.0141$ &  $0.0141$\\

& $b_{11}$  \quad  & $16.00$ & $11.93$  &$11.45$&  $11.15$ \\ \hline

 $\frac{\alpha_{e}(Q^{2})}{\alpha_{s}^{NLO}(Q^{2})}$&$a_{20}$ \quad &$0.0091$ &  $0.010$  &  $0.0124$&  $0.0154$\\

 & $a_{21}$  \quad  &$0.0175$ & $ 0.0186$ & $0.0187$ &  $0.0205$\\

& $b_{21}$  \quad  & $17.12$ & $13.10$  &$12.57$&  $11.64$ \\ \hline
 $\frac{\alpha_{s}^{NLO}(Q^{2})}{4\pi}$&$a_{30}$ \quad &$0.0015$ &  $0.0014$  &  $0.0014$&  $0.0014$\\

 & $a_{31}$  \quad  &$0.0221$ & $ 0.0228$ & $0.0228$ &  $0.0228$\\

& $b_{31}$  \quad  & $15.00$ & $13.01$  &$12.94$&  $12.90$ \\
\bottomrule[1pt]
\end{tabular} }
\endgroup
\end{center}
\end{table}

\newpage
\renewcommand{\theequation}{A-\arabic{equation}}
  \setcounter{equation}{0}  
  \section*{Appendix}  
The coefficients $A_{1}...C_{3}$ in Eqs. (31-33) are as follows
$$
A_{1}(s,u)=\frac{R_{1}(s,u)R{2}(s,u)}{H_{1}(s)},
$$
$$
A_{2}(s,u)=\frac{a_{10}\Omega_{f}(s)}{H_{1}(s,u)},
$$
\begin{equation}
A_{3}(s,u)=\frac{R_{3}(s,u)\Theta_{f}(s)}{H_{1}(s,u)}
\end{equation}

$$
B_{1}(s,u)=\frac{R_{3}\Theta_{g}(s)}{H_{1}(s,u)},
$$
$$
B_{2}(s,u)=\frac{R_{1}(s,u)R_{3}-a_{10}^{2}7\Omega_{f}(s)\Upsilon_{m}(s)}{H_{1}(s,u)},
$$
\begin{equation}
B_{3}(s,u)=\frac{a_{10}\Omega_{f}(s)\Theta_{g}(s)}{H_{1}(s,u)},
\end{equation}

$$
C_{1}(s,u)=\frac{a_{10}R_{2}(s,u)\Upsilon_{m}(s)}{H_{1}(s,u)},
$$
$$
C_{2}(s,u)=\frac{a_{10}\Theta_{f}(s)\Upsilon_{m}(s)}{H_{1}(s,u)},
$$
\begin{equation}
C_{3}(s,u)=\frac{R_{1}(s,u)R_{2}(s,u)-\Theta_{f}\Theta_{g}}{H_{1}(s,u)},
\end{equation}
and coefficients $A_{1}{'}...C_{3}{'}$ in Eqs. (43-45) are as follows
$$
A_{1}{'}(s,u)=\frac{1}{H_{1}(s,u)} \bigg[R_{2}(s,u)R_{3}(s,u)+a_{11}R_{2}(s,u)R_{3}(s,u)\Omega_{f}(s)C_{1}(s,u-b_{11})+a_{10}^{2}R_{2}(s,u)\Omega_{f}(s)$$ $$\times\Upsilon_{m}(s)A_{1}(s,u-b_{11})+a_{11}R_{2}(s,u)R_{3}(s)\Upsilon(s)A_{1}(s,u-b_{11})+a_{10}^{2}R_{2}(s)\Omega_{f}(s)\Omega_{m}(s)C_{1}(s,u-b_{11})\bigg]
$$
$$
A_{2}{'}(s,u)=\frac{1}{H_{1}(s,u)}\bigg[R_{3}(s,u)\Omega_{f}(s)+a_{11}R_{2}(s,u)R_{3}(s)\Upsilon_{f}(s)A_{2}(s,u-b_{11})+a_{10}^{2}R_{2}(s,u)\Omega_{f}(s)$$ $$\times\Upsilon_{m}(s)A_{2}(s,u-b_{11})+a_{11}R_{2}(s,u)R_{3}(s,u)\Omega_{f}(s)C_{2}(s,u-b_{11})+a_{10}^{2}
R_{2}(s)\Omega_{f}\Omega_{m}C_{2}(s,u-b_{11}) \bigg]$$
 $$
A_{3}{'}(s,u)=\frac{1}{H_{1}(s,u)}\bigg[a_{11}R_{2}(s,u)R_{3}(s,u)\upsilon_{f}(s)A_{3}(s,u-b_{11})+a_{10}^{2}R_{2}(s,u)\Omega_{f}(s)\Upsilon_{m}(s)$$
\begin{equation}
\times A_{3}(s,u-b_{11})+a_{11}R_{2}(s,u)R_{3}(s,u)\Omega_{f}(s)C_{3}(s,u-b_{11})+a_{10}^{2}R_{2}(s,u)\Omega_{f}\Omega_{m}C_{3}(s,u-b_{11}) \bigg]
\end{equation}
$$
B_{1}{'}(s,u)=\frac{1}{H_{1}(s,u)}\bigg[R_{3}(s,u)\Theta_{g}(s)+a_{11}R_{3}(s,u)\Theta_{g}(s)\Upsilon{f}(m)A_{1}(s,u-b_{11})+a_{10}^{2}\Omega_{f}(s)\Theta_{g}(s)$$ $$\times\Upsilon_{m}(s)A_{1}(s,u-b_{11})+a_{11}R_{3}(s,u)\Omega_{f}(s)\Theta_{g}(s)C_{1}(s,u-b_{11})+a_{10}^{2}\Omega_{f}(s)\Omega{m}(s)\Theta_{g}C_{1}(u-b_{11})\bigg]
$$

$$
B_{2}{'}(s,u)=\frac{1}{R_{2}(s,u)}+\frac{1}{H_{1}(s,u)}\bigg[\frac{R_{3}(s,u)\Theta_{f}(s)\Theta_{g}(s)}{R_{2}(s,u)}+a_{11}R_{3}(s,u)\Theta_{g}(s)\Upsilon_{f}(s)A_{2}(s,u-b_{11})+$$$$a_{10}^{2}\Omega_{f}(s)\Theta_{g}(s)\Upsilon_{m}(s)A_{2}(s,u-b_{11})+a_{11}R_{3}(s,u)\Omega_{f}(s)\Omega_{m}(s)\Theta_{g}C_{2}(s,u-b_{11})\bigg]
$$

$$
B_{3}{'}(s,u)=\frac{1}{H_{1}(s,u)}\bigg[a_{10}\Omega_{f}(s)\Theta_{g}(s)+a_{11}R_{3}(s,u)\Theta_{g}(s)\Upsilon_{f}(s)A_{3}(s,u-b_{11})+a_{10}^{2}\Omega_{f}(s)\Theta_{g}(s)$$

\begin{equation}
\times \Upsilon_{m}(s)A_{3}(s,u-b_{11})+a_{11}R_{3}(s,u)\Omega_{f}(s)\Theta_{g}(s)C_{3}(s,u-b_{11})+a_{10}^{2}\Omega_{f}(s)\Omega_{m}(s)\Theta_{g}C_{3}(s,u-b_{11})\bigg],
\end{equation}
$$
C{'}_{1}(s,u)=\frac{a_{10}\Upsilon_{m}(s)A_{1}(s,u-b_{11})+a_{10}\Omega_{m}(s)C_{1}(s,u-b_{11})}{R_{3}(s,u)}+\frac{1}{H_{1}(s,u)}\bigg[a_{10}R_{2}(s,u)\Upsilon_{m}(s)+$$ $$a_{10}a_{11}R_{2}(s,u)\Upsilon_{f}(s)\Upsilon_{m}(s)A_{1}(s,u-b_{11})+a_{10}a_{11}R_{2}(s,u)\Omega_{f}(s)\Upsilon_{m}(s)C_{1}(s,u-b_{11})+\frac{1}{R_{3}(s,u)}$$$$\times\left[a_{10}^{3}R_{2}(s,u)\Omega{f}(s)\Upsilon_{m}^{2}(s)A_{1}(s,u-b_{11})+a_{10}^{3}R_{2}(s,u)\Omega_{f}\Omega_{m}\Upsilon{m}
C_{1}(s,u-b_{11})\right]\bigg]$$
$$
 C{'}_{2}(s,u)=\frac{a_{10}\Upsilon_{m}(s)A_{2}(s,u-b_{11})+a_{10}\Omega_{m}(s)C_{2}(s,u-b_{11})}{R_{3}(s,u)}+\frac{1}{H_{1}(s,u)}\bigg[a_{10}\Theta_{f}(s)\Upsilon_{m}(s)+$$ $$a_{10}a_{11}R_{2}(s,u)\Upsilon_{f}(s)\Upsilon_{m}(s)A_{2}(s,u-b_{11})+a_{10}a_{11}R_{2}(s,u)\Omega_{f}(s)\Upsilon_{m}(s)C_{2}(s,u-b_{11})+\frac{1}{R_{3}(s,u)}$$$$\times\left[a_{10}^{3}R_{2}(s,u)\Omega{f}(s)\Upsilon_{m}^{2}(s)A_{2}(s,u-b_{11})+a_{10}^{3}R_{2}(s,u)\Omega_{f}\Omega_{m}\Upsilon{m}
C_{2}(s,u-b_{11})\right]\bigg]
 $$
 $$
 C{'}_{3}(s,u)=\frac{1+a_{10}\Upsilon_{m}(s)A_{3}(s,u-b_{11})+a_{10}\Omega_{m}(s)C_{3}(s,u-b_{11})}{R_{3}(s,u)}+\frac{1}{H_{1}(s,u)}\bigg[a_{10}a_{11}R_{2}(s,u)\Upsilon_{f}(s)$$ $$\times \Upsilon_{m}(s)A_{3}(s,u-b_{11})+a_{10}a_{11}R_{2}(s,u)\Omega_{f}(s)\Upsilon_{m}(s)C_{3}(s,u-b_{11})+\frac{1}{R_{3}(s,u)}\big[a_{10}^{2}R_{2}(s,u)\times$$
\begin{equation}
\Omega_{f}(s)\Upsilon_{m}(s)+a_{10}^{3}R_{2}(s,u)\Omega_{f}(s)\Upsilon_{m}^{2}(s)A_{3}(s,u-b_{11})+a_{10}^{3}R_{2}(s,u)\Omega_{f}\Omega_{m}\Upsilon{m}
C_{3}(s,u-b_{11})\big]\bigg]
\end{equation}
where
$$
H_{1}(s,u)=R_{1}(s,u)R_{2}(s,u)R_{3}(s,u)-R_{3}(s,u)\Theta_{f}(s)\Theta_{g}(s)-a{0}^{2}R_{2}(s,u)\Omega_{f}(s)\Upsilon_{m}(s)
$$
$$
R_{1}(s,u)=u-\Phi_{f}(s)-\Upsilon_{f}(s)a_{10}, \ \ R_{2}(s,u)=u-\Phi_{g}(s), \ \ R_{3}(s,u)=u-\Omega_{m}(s)a_{10}
$$
The coefficients  $A_{1}^{NLO}... C_{3}^{NLO}$ in Eqs. (57-59) are as follow
$$
A_{1}^{NLO}(s,u)=\frac{Y_{2}(s,u)Z_{2}(s,u)}{H_{2}(s,u)}, A_{2}^{NLO}(s,u)=\frac{X_{2}(s)Z_{2}(s,u)}{H_{2}(s,u)},A_{3}^{NLO}(s,u)=\frac{Y_{2}(s,u)X_{3}(s)}{H_{2}(s,u)}
$$
$$
B_{1}^{NLO}(s,u)=\frac{Y_{1}(s)Z_{2}(s,u)}{H_{2}(s,u)},B_{2}^{NLO}(s,u)=\frac{X_{1}(s,u)Z_{2}(s,u)-X_{3}(s)Z^{1}(s)}{H_{2}(s,u)}, B_{3}^{NLO}(s,u)=\frac{X_{3}(s)Y_{1}(s)}{H_{2}(s,u)}
$$
$$
C_{1}^{NLO}(s,u)=\frac{Y_{2}(s,u)Z_{1}(s)}{H_{2}(s,u)},C_{2}^{NLO}(s,u)=\frac{X_{2}(s)Z_{1}(s)-X_{3}(s)Z^{1}(s)}{H_{2}(s,u)}, $$
\begin{equation}
C_{3}^{NLO}(s,u)=\frac{X_{1}(s,u)Y_{2}(s,u)-X_{2}(s)Y_{1}(s)}{H_{2}(s,u)}
\end{equation}
where

$$
H_{2}(s,u)=X_{1}(s,u)Y_{2}(s,u)Z_{2}(s,u)-X_{2}(s)Y_{1}(s)Z_{2}(s,u)-X_{3}(s)Y_{2}(s,u)Z_{1}(s)
$$
$$
X_{1}(s,u)=u-\Phi_{f}^{LO}(s)-a_{30}\Phi_{f}^{NLO}(s)-a_{20}\Upsilon_{f}(s), \ \
X_{2}(s)=\Theta_{f}^{LO}(s)+a_{30}\Theta_{f}^{NLO}(s),\ \
X_{3}(s)=a_{20}\Omega_{f}(s)
$$
$$
Y_{1}(s)=\Theta_{g}^{LO}(s)+a_{30}\Theta_{g}^{NLO}(s)
Y_{2}(s,u)=u-\Phi_{g}^{LO}(s)-a_{30}\Phi_{g}^{NLO}(s)
$$
$$
Z_{1}(s)=a_{20}\Upsilon_{m}(s),\ \ \ Z_{2}(s,u)=u-a_{20}\Omega_{m}(s)
$$
And the coefficients  $A{'}^{NLO}_{1}(s,u)...C{'}^{NLO}_{3}(s,u)$ in Eqs. (63-65) are as follow
$$
A{'}^{NLO}_{1}(s,u)=\frac{1}{H_{2}(s,u)}\Bigg[\bigg[C_{1}^{NLO}(s,u-b_{21})\Omega_{m}(s)+A_{1}^{NLO}(s,u-b_{12})\Upsilon_{m}(s)\bigg]a_{21}X_{3}(s)Y_{2}(s,u)
$$
$$
+\bigg[B_{1}^{NLO}(s,u+b_{31})\Phi_{f}(s)+A_{1}^{NLO}(s,u+b_{31})\Theta_{g}^{NLO}(s)\bigg]a_{31}X_{2}(s)Z_{2}(s,u)
$$
$$
+\bigg[1+a_{21}C_{1}^{NLO}(s,u-b_{21})\Omega_{f}(s)+a_{31}A_{1}^{NLO}(s,u+b_{31})\Phi_{f}^{NLO}(s)+a_{31}B_{1}^{NLO}(s,u+b_{31})\Theta_{f}^{NLO}(s)
$$
$$
+a_{12}A_{1}^{NLO}(s,u-b_{12})\Upsilon_{f}(s)\bigg]Y_{2}(s,u)Z_{2}(s,u)\Bigg]
$$

$$
A{'}^{NLO}_{2}=\frac{1}{H_{2}(s,u)}\Bigg[\bigg[C_{3}^{NLO}(s,u-b_{21})\Omega_{m}(s)+A_{2}^{NLO}(s,u-b_{12})\Upsilon_{m}(s)\bigg]a_{21}X_{3}(s)Y_{2}(s,u)
$$
$$
+\bigg[B_{2}^{NLO}(s,u+b_{31})\Phi_{f}(s)+A_{2}^{NLO}(s,u+b_{31})\Theta_{g}^{NLO}(s)\bigg]a_{31}X_{2}(s)Z_{2}(s,u)
$$
$$
+\bigg[1+a_{21}C_{3}^{NLO}(s,u-b_{21})\Omega_{f}(s)+a_{31}A_{2}^{NLO}(s,u+b_{31})\Phi_{f}^{NLO}(s)+a_{31}B_{2}^{NLO}(s,u+b_{31})\Theta_{f}^{NLO}(s)
$$
$$
+a_{12}A_{2}^{NLO}(s,u-b_{12})\Upsilon_{f}(s)\bigg]Y_{2}(s,u)Z_{2}(s,u)\Bigg]
$$

$$
A{'}^{NLO}_{3}=\frac{1}{H_{2}(s,u)}\Bigg[\bigg[C_{3}^{NLO}(s,u-b_{21})\Omega_{m}(s)+A_{3}^{NLO}(s,u-b_{12})\Upsilon_{m}(s)\bigg]a_{21}X_{3}(s)Y_{2}(s,u)
$$
$$
+X_{3}(s)Y_{2}(s,u)+\bigg[B_{3}^{NLO}(s,u+b_{31})\Phi_{f}(s)+A_{3}^{NLO}(s,u+b_{31})\Theta_{g}^{NLO}(s)\bigg]a_{31}X_{2}(s)Z_{2}(s,u)
$$
$$
+\bigg[a_{21}C_{3}^{NLO}(s,u-b_{21})\Omega_{f}(s)+a_{31}A_{3}^{NLO}(s,u+b_{31})\Phi_{f}^{NLO}(s)+a_{31}B_{3}^{NLO}(s,u+b_{31})\Theta_{f}^{NLO}(s)
$$
\begin{equation}
+a_{12}A_{3}^{NLO}(s,u-b_{12})\Upsilon_{f}(s)\bigg]Y_{2}(s,u)Z_{2}(s,u)\Bigg]
\end{equation}
$$
B{'}_{1}^{NLO}(s)=\frac{a_{31}}{Y_{2}(s,u)}\bigg[B_{1}^{NLO}(s,u+b_{31})\Phi_{f}^{NLO}(s)+A_{1}^{NLO}(s,u+b_{31})\Theta_{g}^{NLO}(s)\bigg]
$$
$$
+\frac{1}{H_{2}(s,u)}\Bigg[\bigg[C_{1}^{NLO}(s,u-b_{21})\Omega_{m}(s)+A_{1}^{NLO}(s,u-b_{21})\Upsilon_{m}(s)\bigg]a_{21}X_{3}Y_{1}(s)+\bigg[1+a_{21}C_{1}^{NLO}(s,u-b_{12})\Omega_{f}(s)
$$
$$
+a_{31}A_{1}^{NLO}(s,u+b_{31})\Phi_{f}^{NLO}(s)+a_{31}B_{1}^{NLO}(s,u+b_{31})\Theta_{f}^{NLO}(s)
$$
$$
+a_{21}A_{1}^{NLO}(s,u-b_{21})\Upsilon_{f}(s)+\frac{X_{2}(s)}{Y_{2}(s,u)}\bigg[a_{31}B_{1}^{NO}(s,u+b_{31})\Phi_{f}^{NLO}(s)+a_{31}A_{1}^{NLO}(s,u+b_{31})\Theta_{g}(s)\bigg]\bigg]Y_{1}(s)Z_{2}(s,u)\Bigg]
$$
$$
B{'}_{2}^{NLO}(s)=\frac{1}{Y_{2}(s,u)}\bigg[1+a_{31}B_{2}^{NLO}(s,u+b_{31})\Phi_{f}^{NLO}(s)+a_{31}A_{2}^{NLO}(s,u+b_{31})\Theta_{g}^{NLO}(s)\bigg]
$$
$$
+\frac{1}{H_{2}(s,u)}\Bigg[\bigg[C_{3}^{NLO}(s,u-b_{21})\Omega_{m}(s)+A_{2}^{NLO}(s,u-b_{21})\Upsilon_{m}(s)\bigg]a_{21}X_{3}(s)Y_{1}(s)+\bigg[a_{21}C_{3}^{NLO}(s,u-b_{12})\Omega_{f}(s)
$$
$$
+a_{31}A_{2}^{NLO}(s,u+b_{31})\Phi_{f}^{NLO}(s)+a_{31}B_{2}^{NLO}(s,u+b_{31})\Theta_{f}^{NLO}(s)
$$
$$
+a_{21}A_{2}^{NLO}(s,u-b_{21})\Upsilon_{f}(s)+\frac{X_{2}(s)}{Y_{2}(s,u)}\bigg[1+a_{31}B_{1}^{NO}(s,u+b_{31})\Phi_{f}^{NLO}(s)+a_{31}A_{1}^{NLO}(s,u+b_{31})\Theta_{g}(s)\bigg]\bigg]Y_{1}(s)Z_{2}(s,u)\Bigg]
$$
$$
B{'}_{3}^{NLO}(s)=\frac{a_{31}}{Y_{2}(s,u)}\bigg[B_{3}^{NLO}(s,u+b_{31})\Phi_{f}^{NLO}(s)+A_{3}^{NLO}(s,u+b_{31})\Theta_{g}^{NLO}(s)\bigg]
$$
$$
+\frac{1}{H_{2}(s,u)}\Bigg[\bigg[1+C_{3}^{NLO}(s,u-b_{21})\Omega_{m}(s)+A_{3}^{NLO}(s,u-b_{21})\Upsilon_{m}(s)\bigg]a_{21}X_{3}(s)Y_{1}(s)+\bigg[a_{21}C_{3}^{NLO}(s,u-b_{12})\Omega_{f}(s)
$$
$$
+a_{31}A_{3}^{NLO}(s,u+b_{31})\Phi_{f}^{NLO}(s)+a_{31}B_{3}^{NLO}(s,u+b_{31})\Theta_{f}^{NLO}(s)
$$
\begin{equation}
+a_{21}A_{3}^{NLO}(s,u-b_{21})\Upsilon_{f}(s)+\frac{X_{2}(s)}{Y_{2}(s,u)}\bigg[a_{31}B_{3}^{NO}(s,u+b_{31})\Phi_{f}^{NLO}(s)+a_{31}A_{3}^{NLO}(s,u+b_{31})\Theta_{g}(s)\bigg]\bigg]Y_{1}(s)Z_{2}(s,u)\Bigg]
\end{equation}
$$
C{'}_{1}^{NLO}(s)=\frac{a_{21}}{Z_{2}(s,u)}\bigg[C_{1}^{NLO}(s,u-b_{21})\Omega_{m}(s)+A_{1}^{NLO}(s,u-b_{21})\Upsilon_{m}(s)\bigg]
$$
$$
+\frac{1}{H_{2}(s,u)}\Bigg[1+\bigg[B_{1}^{NLO}(s,u+b_{31})\Phi_{f}^{NLO}(s)+A_{1}^{NLO}(s,u+b_{31})\Theta_{g}(s)\bigg]a_{31}X_{2}(s)Z_{1}(s)+\bigg[a_{21}C_{1}^{NLO}(s,u-b_{12})\Omega_{f}(s)
$$
$$
+a_{31}A_{1}^{NLO}(s,u+b_{31})\Phi_{f}^{NLO}(s)+a_{31}B_{1}^{NLO}(s,u+b_{31})\Theta_{f}^{NLO}(s)
$$
$$
+a_{21}A_{1}^{NLO}(s,u-b_{21})\Upsilon_{f}(s)+\frac{X_{3}(s)}{Z_{2}(s,u)}\bigg[a_{21}C_{1}^{NO}(s,u-b_{21})\Omega_{m}(s)+a_{21}A_{1}^{NLO}(s,u-b_{21})\Upsilon_{m}(s)\bigg]\bigg]Y_{2}(s,u)Z_{1}(s)\Bigg]
$$
$$
C{'}_{2}^{NLO}(s)=\frac{a_{21}}{Z_{2}(s,u)}\bigg[C_{3}^{NLO}(s,u-b_{21})\Omega_{m}(s)+A_{2}^{NLO}(s,u-b_{21})\Upsilon_{m}(s)\bigg]
$$
$$
+\frac{1}{H_{2}(s,u)}\Bigg[1+a_{31}\bigg[B_{2}^{NLO}(s,u+b_{31})\Phi_{f}^{NLO}(s)+a_{31}A_{2}^{NLO}(s,u+b_{31})\Theta_{g}(s)\bigg]X_{2}(s)Z_{1}(s)+\bigg[a_{21}C_{3}^{NLO}(s,u-b_{12})\Omega_{f}(s)
$$
$$
+a_{31}A_{2}^{NLO}(s,u+b_{31})\Phi_{f}^{NLO}(s)+a_{31}B_{2}^{NLO}(s,u+b_{31})\Theta_{f}^{NLO}(s)
$$
$$
+a_{21}A_{2}^{NLO}(s,u-b_{21})\Upsilon_{f}(s)+\frac{X_{3}(s)}{Z_{2}(s,u)}\bigg[a_{21}C_{3}^{NO}(s,u-b_{21})\Omega_{m}(s)+a_{21}A_{2}^{NLO}(s,u-b_{21})\Upsilon_{m}(s)\bigg]\bigg]Y_{2}(s,u)Z_{1}(s)\Bigg]
$$
$$
C{'}_{3}^{NLO}(s)=\frac{1}{Z_{2}(s,u)}\bigg[1+a_{21}C_{3}^{NLO}(s,u-b_{21})\Omega_{m}(s)+a_{21}A_{3}^{NLO}(s,u-b_{21})\Upsilon_{m}(s)\bigg]
$$
$$
+\frac{1}{H_{2}(s,u)}\Bigg[a_{31}\bigg[B_{3}^{NLO}(s,u+b_{31})\Phi_{f}^{NLO}(s)+a_{31}A_{3}^{NLO}(s,u+b_{31})\Theta_{g}(s)\bigg]X_{2}(s)Z_{1}(s)+\bigg[a_{21}C_{3}^{NLO}(s,u-b_{12})\Omega_{f}(s)
$$
$$
+a_{31}A_{3}^{NLO}(s,u+_{31})\Phi_{f}^{NLO}(s)+a_{31}B_{3}^{NLO}(s,u+b_{31})\Theta_{f}^{NLO}(s)
$$
\begin{equation}
+a_{21}A_{3}^{NLO}(s,u-b_{21})\Upsilon_{f}(s)+\frac{X_{3}(s)}{Z_{2}(s,u)}\bigg[1+a_{21}C_{3}^{NO}(s,u-b_{21})\Omega_{m}(s)+a_{21}A_{3}^{NLO}(s,u-b_{21})\Upsilon_{m}(s)\bigg]\bigg]Y_{2}(s,u)Z_{1}(s)\Bigg]
\end{equation}

\end{document}